%% file: base_main.tex
\documentclass[aps,prb,twocolumn]{revtex4}

\usepackage{graphicx}

\begin{document}
\title{On velocity profiles, stresses and Bagnold scaling of sheared
  granular system in zero gravity}

\newcommand{\av}[1]{\langle #1 \rangle}
\newcommand \fourtitles {
\hbox{\scriptsize
\hspace{0.5in} without oscillations
\hspace{0.5in} without oscillations
\hspace{0.7in} with oscillations  
\hspace{0.7in} with oscillations
}
\hbox{\scriptsize
\hspace{0.5in} with glued particles
\hspace{0.4in} without glued particles
\hspace{0.5in} with glued particles
\hspace{0.4in} without glued particles
}}
\newcommand \abcd {\vspace{-0.05in} \hbox{\hspace{0.8in} (a)
    \hspace{1.4in} (b) \hspace{1.4in} (c) \hspace{1.4in} (d) } }

\author{Oleh Baran}
\email[]{oleh.baran@njit.edu}
\homepage[]{http://math.njit.edu/~oleh/}
\affiliation{Department of Mathematical Sciences and Center for
  Applied Mathematics and Statistics, New Jersey Institute
of Technology, Newark, NJ 07102, USA}

\author{Lou Kondic}
\affiliation{Department of Mathematical Sciences and Center for
  Applied Mathematics and Statistics, New Jersey Institute
of Technology, Newark, NJ 07102, USA}

\date{August 1, 2004}
\begin{abstract}
We report the results of three-dimensional molecular dynamics
simulations of sheared granular system in a Couette
geometry.\cite{movies} The simulations use realistic boundary
conditions that may be expected in physical experiments. For a range
of boundary properties we report velocity and density profiles, as
well as forces on the boundaries. In particular, we find that the
results for the velocity profiles throughout the shearing cell depend
strongly on the interaction of the system particles with the physical
boundaries.  Even frictional boundaries can allow for significant
slippage of the particles, therefore, reducing the shear in the
system. Next, we present shear rate dependence of stress, including
mean force and force fluctuations, both for controlled volume, and for
controlled stress configurations. We discuss the dependence of solid
volume fraction on shear rate under the constant pressure condition,
and Bagnold scaling in volume controlled experiments. In addition, we
study the influence of oscillatory driving on the system properties.
\end{abstract}
\begin{flushright} 
\end{flushright}

\maketitle

\input{base_intro}  
\input{base_model}

\input{base_glued}
\input{base_noshaking}
\input{base_shaking}
\input{base_forces}
\input{base_conclusions}

\appendix


\bibliographystyle{unsrt}
\bibliography{p1_2004.bib}

\end{document}

%% file: base_intro.tex
\section{Introduction}

A significant understanding of granular flow results from experimental
measurements and numerical simulations. Simulations have proved to
effectively complement the experimental measurements. Among many known
advantages of computer simulations in any field of physics we can
distinguish the following three types: 1) The ability to study the
regions of parameter phase space that are difficult to access in
experiment. For example, regarding granular experiments,
high-frequency and/or high amplitude driving can be easily simulated
but is difficult to achieve in experiments because of considerable
power required; 2) The ability to study the system under (often
simplified) conditions that are not reachable or are very difficult to
have in experiments. For example, 2D simulations as opposed to 3D
experiments, periodic boundary conditions as opposed to solid walls
boundaries, etc.; 3) The ability to obtain more complete and detailed
information under the same or close to the same conditions as in an
experimental study.

However, in the subfield of sheared granular flow, most of the known
numerical studies exploit either the first or the second advantage of
simulations from the above list.
\cite{aharonov02,campbell85,campbell89,silbert01,campbell02,alam03,alam03b,alam02}
In particular, there are very few efforts to model realistically the
physical boundaries, and this is the most common discrepancy between
the set-up of laboratory experiments and the set-up of numerical
simulations. This effect of physical boundaries, such as the walls of
the container, is often and rightfully considered responsible for the
discrepancy between the results of the experiments and of the
simulations, or between the predictions of the existing theories and
the results of experiments or simulations.\cite{alam98,nott99} Jenkins
and Richman\cite{jenkins86}, for example, calculated boundary
conditions in a specific limit for plane flow of identical, smooth,
inelastic disks interacting with a bumpy wall. Louge, Jenkins and
Hopkins\cite{louge90} and later Louge\cite{louge94} tested these
theoretical predictions for rapid sheared granular flows using
computer simulations.
The work of Nott, Alam, Agrawal, Jackson and Sundaresan{\em
et~al.}\cite{nott99} presents the theoretical study of the effect of
boundaries on the plane Couette flow, indicating the possibility of
many different stable and unstable states of the flow, completely
determined by the properties of the boundaries.
However, to the best of our knowledge, none of the simulations
reported in the literature for sheared granular flow are using the
boundary models that closely reflect the geometry and the conditions
of a Couette cell. Instead, simple models and boundary conditions are
conventionally used to set and study sheared granular flow. Campbell
and Brennen\cite{campbell85}, for example, recognize the effect of
shearing wall properties on velocity and granular temperature profile
in 2D shearing cell by presenting in parallel the results for two
models of particle-wall collisions: Type A model assumes the
interaction between particles and the wall is similar to the
interaction between particles and particles, and; Type B model
attempts to approximate a no-slip condition by setting the
after-rebounce velocity of a colliding particle equal to the velocity
of the wall. Type B model has then been studied more extensively by
many authors because it provides a constant shear rate and constant
volume fraction throughout the sample and thus makes a convenient test
case when comparison with theoretical predictions is
considered. However, the experimental data indicate strongly the
presence of boundary effects on the shear flow. The examples are the
formation of a shear band, or other non-linear shear rate and volume
fraction distributions in the sample (see
e.g. Ref.~\onlinecite{losert00}), or the drastic difference between
the shearing using a wall with glued particles on it as opposed to
shearing the system by a plane frictional wall. Similarly, the
periodic boundary conditions, that are often used in simulations,
neglect the effects of stationary walls on the shear flow.


In this paper we report the systematic study of the effect of boundary
conditions on sheared granular flow in a Couette cell in zero gravity
using event-driven simulations. The paper is organized as
follows. Section~\ref{sec:ed} describes the numerical approach and
relevant parameters of simulations. In Sec.~\ref{sec:shear} we present
the results for bulk velocity and volume fraction profiles for the
variety of boundary conditions, such as the system with rough shearing
wall with and without glued particles on it. Also we consider
simulations with oscillating bottom wall. We show the results for the
systems of varied sizes, varied properties of the walls of a Couette
cell, varied intensity of external driving, and discuss the effect of
each varied parameter on the profiles and equilibrating times. In
Sec.~\ref{sec:forces} we discuss stresses on the physical boundaries
of a Couette cell and their distributions. We compare the normal and
shear component of the stress distributions and discuss the factors
determining the widths of the distributions and their average values.

All the results in Secs.~\ref{sec:gb}-\ref{sec:forces} are obtained
for the systems of fixed or directly controlled total volume. In
Sec.~\ref{sec:cp} we present the results for the stresses, velocities
and volume fraction profiles for the systems with stress-controlled
boundaries. We then discuss the differences between volume controlled
and stress-controlled simulations. In particular, we consider the
different response to the increase of the shearing velocity in these
two configurations.

%% file: base_model.tex
\section{\label{sec:ed} Simulation details}


Numerical algorithm of our choice, event-driven algorithm for hard
spheres, is well described elsewhere, for example in
Ref.~\onlinecite{lubachevsky92}. In zero gravity all particles follow
linear trajectories between the collisions. All collisions between
particles are instantaneous and binary. Consider two colliding
particles with diameter $\sigma_1$ and $\sigma_2$, masses $m_1$ and
$m_2$, positions ${\bf r}_1$ and ${\bf r}_2$, linear and angular
velocities ${\bf v}_1$, ${\bf \omega}_1$ and ${\bf v}_2$, ${\bf
\omega}_2$. Then velocities after collision are:
\begin{eqnarray}
{\bf v}_1^{\prime} = {\bf v}_1 - \frac{m_2}{m_1+m_2} \Delta {\bf V}
\mbox{;} \hspace{0.1cm}
{\bf v}_2^{\prime} = {\bf v}_2 + \frac{m_1}{m_1+m_2} \Delta {\bf V} \\
{\bf \omega}_1^{\prime} = {\bf \omega}_1 +
\frac{m_2}{m_1+m_2} \frac{\Delta {\bf W}}{\sigma_1}
\mbox{;} \hspace{0.1cm}
{\bf \omega}_2^{\prime} = {\bf \omega}_2 +
\frac{m_1}{m_1+m_2} \frac{\Delta {\bf W}}{\sigma_2}
\end{eqnarray}
where
\begin{equation}
\Delta {\bf V} = (1+e){\bf v}_n + \frac{2}{7}(1+\beta) {\bf v}_t
\mbox{;} \hspace{0.1cm}
\Delta {\bf W} =\frac{10}{7} (1+\beta)\, [ {\bf n} \times
{\bf v}_t ]
\end{equation}
Here ${\bf n}=({\bf r}_2 - {\bf r}_1)/|{\bf r}_2 - {\bf r}_1|$ is the
normal unit vector, ${\bf v}_n$ and ${\bf v}_t$ are the normal and
tangential components of the relative velocity ${\bf v}_c= {\bf v}_1 -
{\bf v}_2 - [(\frac{1}{2}\sigma_1 \, {\bf \omega}_1 +
\frac{1}{2}\sigma_2 \, {\bf \omega}_2) \times {\bf n}]$ of particles
at the contact point. Total linear and angular momentum are conserved
during a collision, however, total translational and rotational
energies are lost.

Energy dissipation is controlled by three parameters:\cite{foerster94}
the coefficient of restitution $e$, coefficient of friction $\mu$, and
coefficient of tangential restitution $\beta$. The algorithm is
adapted to include the particle's interaction with physical boundaries
in the following way. The collisions between particles and the top or
bottom walls (also called horizontal walls) assume that the wall is a
particle of infinite radius, infinite mass, and of linear velocity
equal to the wall's velocity at the point of contact. Three
dissipation parameters, $e_w$ $\mu_w$ and $\beta_w$, characterize
these collisions.
In the collisions between free and glued (to be explained below)
particles we use the same dissipation parameters as in the collisions
between free particles. In the calculation of the re-bounce velocity
we assume that a participating glued particle has an infinite
mass. Finally, the side walls are the particles of infinite mass and
of infinite radius and with the surface normal at the point of contact
pointing horizontally toward the center of the cell. The properties of
the side-walls are characterized by the separate set of dissipation
parameters, $e_s$, $\mu_s$ and $\beta_s$.

Experiments and theoretical studies show that the coefficient of
restitution $e$ noticeably depends on the impact
velocity.\cite{goldsmith64,schwager98} To account for this effect, we
set this coefficient for all types of collisions to be velocity
dependent in a way suggested in Ref.~\onlinecite{bizon98}
\begin{equation}
e(v_n) = \left\{ \begin{array}{ll}
1-Bv_n^{3/4}, & v_n<v_0 \\
\epsilon,       & v_n>v_0
\end{array}
\right.
\label{eq:restitution}
\end{equation}
Here $v_n$ is the component of relative velocity along the line
joining particle centers, $B=(1-\epsilon )v_0^{(-3/4 )}$, $v_0 \simeq
100 \, \av{d}/sec$ 
$\epsilon$ is a restitution parameter that characterizes the material
 properties of the particles or walls: Decreasing $\epsilon$ is
 equivalent to increasing the energy dissipated during each collision.

Coefficient of tangential restitution, $\beta$, defined as the ratio
of the tangential components of the relative velocity after and before
collision, is assumed to be the one suggested in Ref.~\onlinecite{walton93}
and used in Ref.~\onlinecite{bizon98}, {\em i.e.}
\begin{equation}
\beta = \left\{ \begin{array}{ll}
-1+ \mu (1+e) (7/2) v_n  / v_t, & $ for sliding contacts$ \\
\beta_0,       & $ for rolling contacts$
\end{array}
\right.
\label{eq:beta}
\end{equation}
Here $\mu$ is the coefficient of friction, $e$ is given by
(\ref{eq:restitution}), and $\beta_0$ is a parameter which also has a
meaning of restitution of velocity in the tangential direction for
sliding contacts. As in Ref.~\onlinecite{walton93}, the choice between
rolling and sliding solution depends on the ratio of normal to
tangential component of relative velocity, and on the parameters
$\beta_0$, $\epsilon$, $\mu$. The same functional dependence is
assumed for $\beta_{w}$, $e_w$, and for $\beta_{s}$, $e_s$.

\begin{table*}[!ht] 
\caption{\label{tab:parameters} Typical choice of dissipation
parameters.}
\begin{ruledtabular}
\begin{tabular}{|c||c|c|c||c|}
~particles (including glued)~& $\epsilon = 0.6$~ & $\mu = 0.5$~ &
$\beta_0 = 0.35$~ & ~frictional inelastic particles~ \\ \hline ~top
and bottom walls~ & $\epsilon_w = 0.1$~ & $\mu_w = 0.9$~ & $\beta_{0w}
= 0.35$~ & ~very frictional and inelastic~ \\ \hline ~side-walls
~ & $\epsilon_s = 0.9$~ & $\mu_s = 0.1$~ & $\beta_{0s} = 0.35$~ &
~slightly frictional and inelastic~ \\
\end{tabular}
\end{ruledtabular}
\end{table*}

Table \ref{tab:parameters} summarizes our choice of dissipation
parameters for typical configurations. When different dissipation
parameters are used in the simulations, the results are always
compared to the typical configuration. 

\begin{figure*}[!ht]
\centerline{\includegraphics{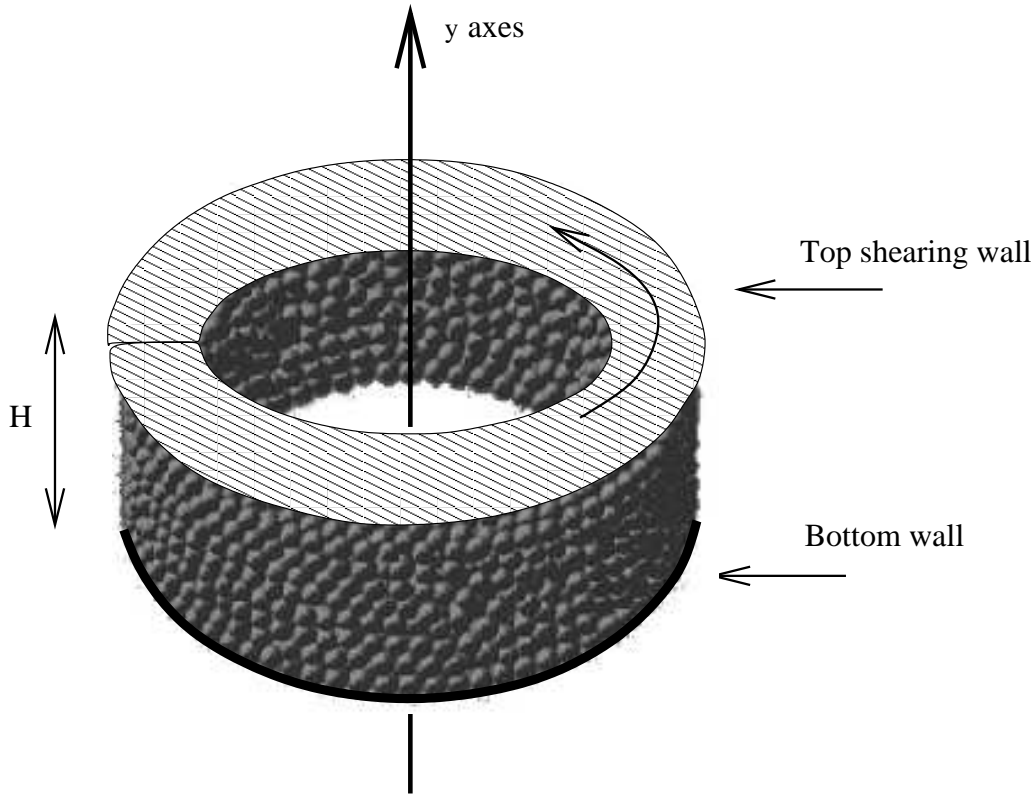}}
\caption{\label{fig:setup1} Granular matter between two concentric
stationary cylinders - side walls (not shown in the figure). Top wall
is rotating around the vertical $y$ axes. Bottom wall can be vibrated
in the vertical direction in volume controlled simulations, or it can
move in the vertical direction in stress controlled simulations.}
\end{figure*}
We have carried out the simulations using a 3D granular system of
polydisperse inelastic frictional spheres in the Couette cell,
Fig.~\ref{fig:setup1}, with stationary side walls, rotated top wall,
and the bottom wall that can move in the vertical direction.

The last feature allows us to oscillate the bottom wall. The
motivation for these oscillations is that they are often used in
laboratory experiments performed with granular systems of high volume
fraction in order to fluidize otherwise jammed
state\cite{daniels03}. We use the oscillating bottom wall to study the
effect of oscillations on the granular flow and to make the connection
with the laboratory experiments, where possible.

In addition, the moving bottom wall feature is actively used in stress
controlled simulations, where we allow granular particles to determine
the position and the velocity of the bottom wall on their own. Here
the stress constraint determines the average volume fraction. We refer
to Section~\ref{sec:cp} for more detailed discussion of this
configuration.

The particles are polydisperse with diameter values randomly
distributed in the range [0.9,1.1] $\av{d}$, where $\av{d}$ is the
average diameter of particles, that is used as a natural length
scale. Our {\it typical configuration} consist of $N=2000$ particles,
and fixed radii of inner cylinder and outer cylinder ($R_i=8 \av{d}$,
$R_o=12\av{d}$). The height $H$ of the Couette cell is variable and in
general is a function of time and preset average total volume fraction
$\nu$
for the volume controlled simulations, and it is the function of time
and preset average stress on the bottom wall in stress controlled
simulations. In the case of volume controlled simulations we have
\begin{equation}
  H(\nu,t) = \left ( 4.22 \frac{1}{\nu} + A sin ( \omega t ) \right
  )\, \av{d}.
  \label{eq:h1}
\end{equation}
Here $A$, $\omega$ are the amplitude scaled by $\av{d}$, and the
frequency in $rad/sec$ of the bottom wall vibrations, if any. Thus, in
absence of vibrations, $H=10.54 \av{d}$ for $\nu=40\%$; $H=21.10
\av{d}$ for $\nu=20\%$ and $H=8.44 \av{d}$ for $\nu=50\%$. Our typical
choices for the amplitude and frequency of the vibrations of the
bottom wall are $A=1 \av{d}$ and $\omega_t = 230 \, rad/sec$ or
$f=\omega_t / 2 \pi = 36.6 \, Hz$. These amplitude and frequency set
the dimensionless parameter $\Gamma = A\omega^2/g$ to the value which
is above critical for the onset of fluidization, $\Gamma =5 >
\Gamma_c=1$. We use the value of gravitational constant $g=9810
\av{d}/sec^2$ ($=981\, cm/sec^2$ when $\av{d}=0.1 \, cm$).

Another system size studied in this work is a {\it wide configuration}
that has $N=5000$ particles in a Couette cell with radii of inner and
outer cylinders $R_i=5 \av{d}$ and $R_o=15 \av{d}$, respectively, and
$H$ given by (\ref{eq:h1}).

\begin{figure*}[!ht]
\centerline{\hbox{
\includegraphics{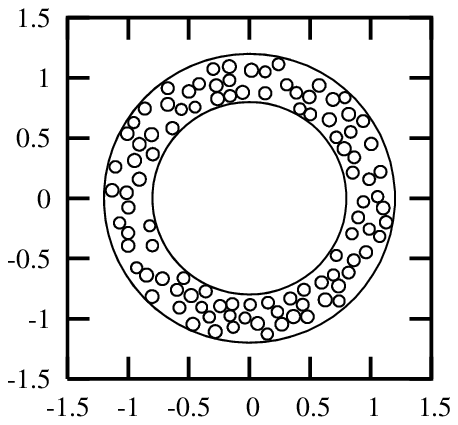}
\includegraphics{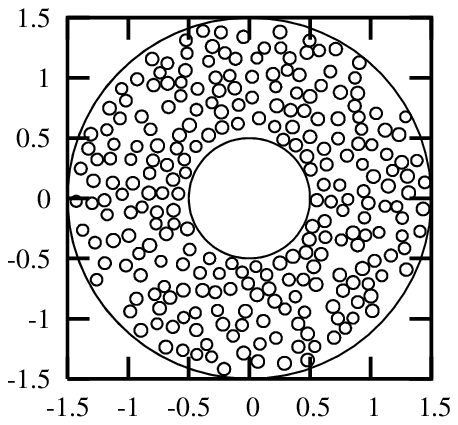}
}}
\centerline{\hbox{\hspace{0.1in} (a) \hspace{1.9in} (b)}}
\caption{\label{fig:gb_conf} (a) Typical configuration: distribution
of 100 glued particles on a shearing surface, surface volume fraction
$\approx$ 30\%; (b) Wide configuration: distribution of 250 glued
particles, surface volume fraction $\approx$ 30\%.}
\end{figure*}

We assume the surface properties of side walls to be the same for
inner and outer cylinder and different from the properties of top and
bottom walls. The top wall can be covered with particles glued to its
surface. These glued particles have the same polydispersity, mean
diameter and material properties as the free particles. They are glued
into the dimples on the surface so that each one protrudes
inside the cell by the distance of half it's
diameter. Figure~\ref{fig:gb_conf} shows a typical distribution of
glued particles on the inside surface of the top wall. For both
typical and wide configurations the number of glued particles
corresponds to $\approx 30\%$ surface fraction.

%% file: base_glued.tex
\section {\label{sec:shear} Velocity and volume fraction profiles}
\begin{figure*}[!ht]
\centerline{\hbox{
\includegraphics{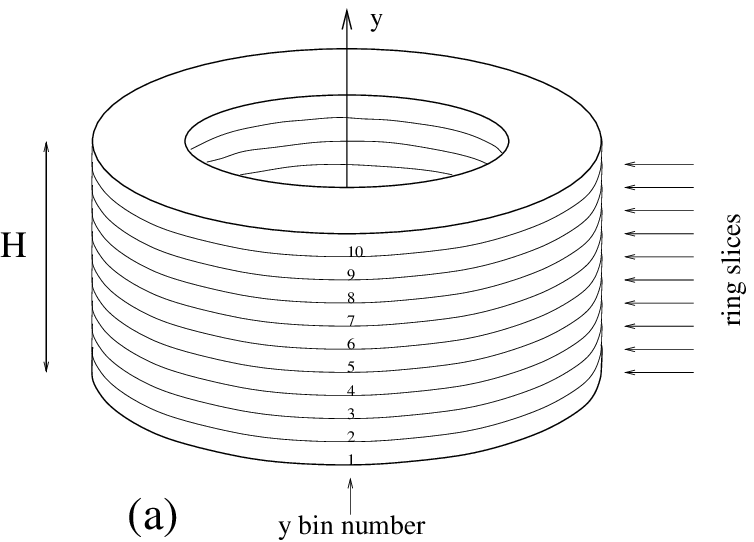}
\includegraphics{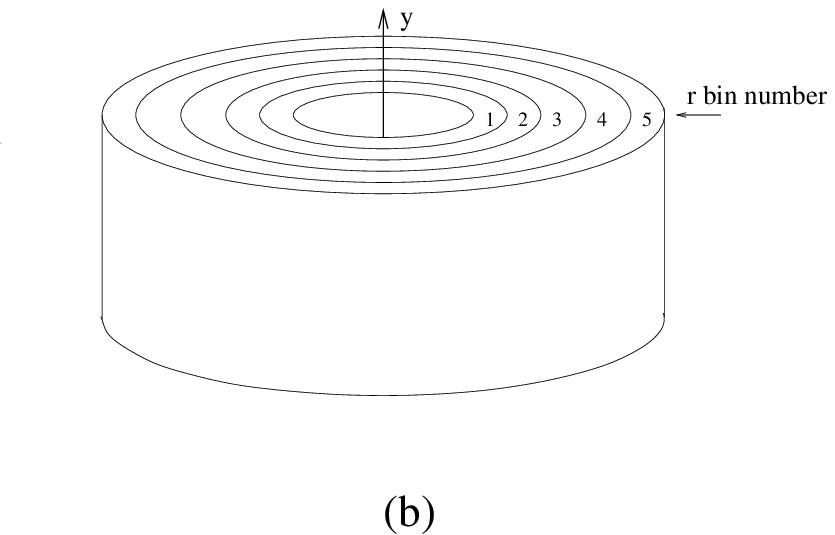}
}}
\caption{\label{fig:ybin} (a) $y$ bin number in measurements of
velocity and volume fraction profiles. (b) $r$ bin number for wide
configurations.}
\end{figure*}

To measure velocity and volume fraction profiles, the volume of
Couette cell is divided in $N_s$ ring slices, Fig. \ref{fig:ybin}
(a). Each slice is assigned a $y$ bin number $i_y=1,...,N_s$. In most
cases, we set $N_s=10$ and the height of one ring slice is $h_y=1.05
\av{d}$. When the bottom wall is oscillated, we account for variable
volume of the bottom slice or slices.
In our measurements we average a studied quantity both over the volume
of a $y$ bin and over the time interval $t_{p}$, typically $t_{p}= 0.5
\, sec$. Within this time interval we measure the particle's
velocities and volume fractions periodically every $\delta t_p$
leading to $N_p=t_{p}/\delta t_p$ different distributions. It is
convenient to express times in terms of the following time scale
\begin{equation}
T_w=\frac{2 \pi}{\omega_t} 
\end{equation}
Time $T_w$ correspond to the period of oscillation with frequency
$\omega_t$.
Thus $t_{p}=18.3 \, T_w$, and we use $\delta t_p=T_w/10$. $T_w$ is
used as a scale independently of whether the system is vibrated or
not. All particles are binned in the corresponding ring slices, and
the average solid volume fraction and the average tangential velocity
in each bin are calculated as a function of $y$ bin number.

For wide configurations, in addition we measure the volume fraction as
a function of the distance to the inner cylinder. In this case the
volume of Couette cell is divided into a set of vertical shells as in
Fig.~\ref{fig:ybin}(b). Each shell is assigned $r$ bin number. The
averaging procedure is analogous to the one described above for $y$
profiles.

In the figures of this paper, the steady state velocity profiles are
shown as lines with filled triangles. The ``shadow of broken lines''
below the steady state velocity profile are transient state velocity
profiles measured periodically - typically every 1/12 of the total
simulation time. Filled circles mark the steady state volume fraction
profiles. Transient states for volume fraction are also shown as the
collection of broken lines.  For each case we show the scaled energy
plot as a function of time. The scaled energy is measured kinetic
energy per particle scaled by the energy of a particle of average
diameter and of average linear top wall velocity. These energy plots
allow to determine whether and when the system reaches a steady state.

\subsection{\label{sec:gb} Typical configurations with inelastic and frictional sidewalls}
\begin{figure*}[!ht]
\fourtitles
\hbox{
\includegraphics{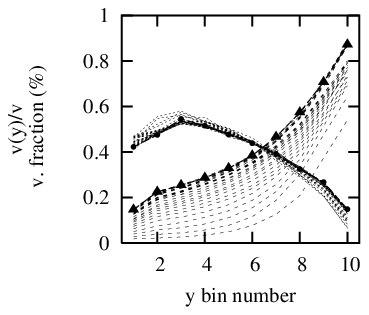} 
\includegraphics{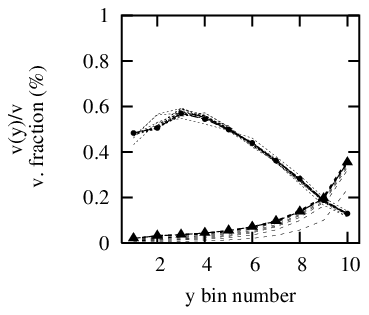} 
\includegraphics{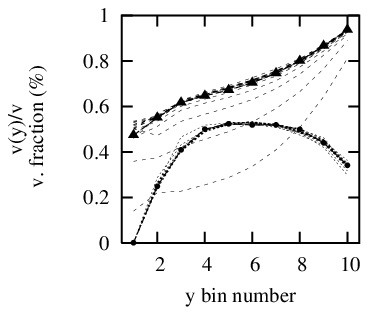} 
\includegraphics{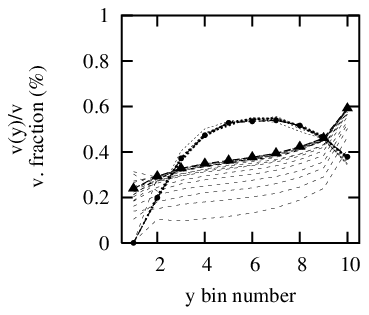} 
}
\hbox{
\includegraphics{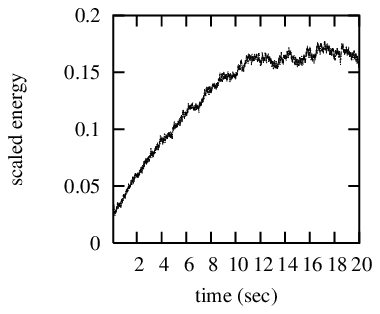}
\includegraphics{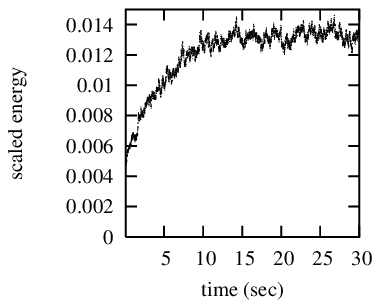}
\includegraphics{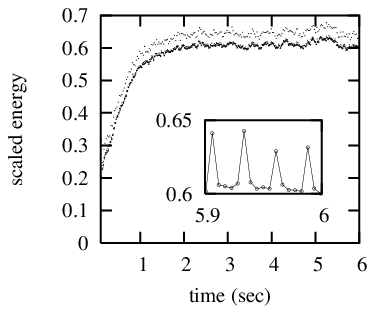}
\includegraphics{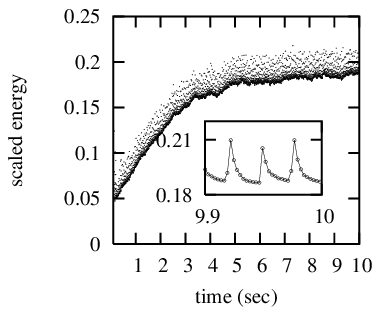}
}
\hbox{
\includegraphics{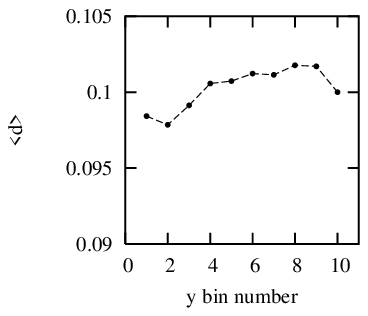}
\includegraphics{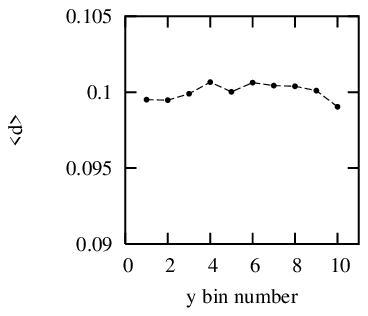}
\includegraphics{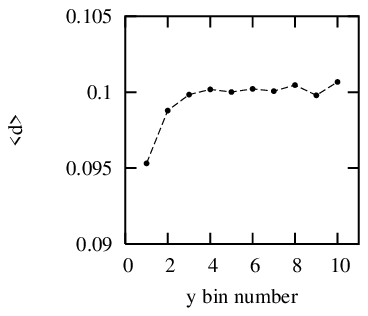}
\includegraphics{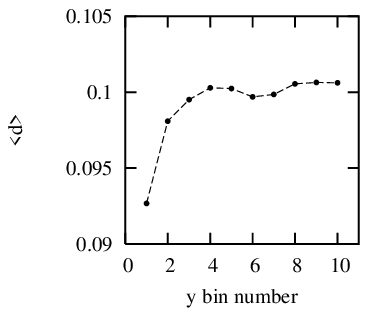}
}
\abcd
\caption{\label{fig:gb} Typical configurations results. First Panel:
scaled velocity profiles (filled triangles) and volume fraction
profiles (filled circles) in steady state regime together with the
profiles in transient regimes (dashed lines). Second panel: scaled
energy plots. Insets in energy plots: kinetic energy plotted on a
shorter time scale showing the frequency of bottom wall
oscillations. Third panel: average diameter of particles. (a), (b) -
results without oscillating bottom wall; (c), (d) - with oscillations;
(a), (c) - shearing wall with 100 glued particles; (b), (d) - without
glued particles.  In all simulations the dissipation parameters are as
shown in Table \ref{tab:parameters}.}
\end{figure*}

Figure~\ref{fig:gb} shows the velocity profiles, top horizontal panel,
scaled total kinetic energy, second horizontal panel, and average size
of particles $\av{d}$, third horizontal panel. All the measurements
are done for the typical configuration systems with 2000 particles and
in zero gravity, but with four different boundary conditions. The
first column of graphs, (a), shows the results for the system with
oscillating base and glued particles on the top wall. The second
column, (b), shows the results for the system with oscillating bottom
wall but without glued particles. The last two columns show the
results for the system without oscillating bottom wall with (c) and
without (d) glued particles. When oscillations are present, the
vertical $y$ coordinate of the bottom wall is given by $y=Asin(\omega
t)$, with amplitude $A=\av{d}$ and frequency $\omega = \omega_t = 230
\, rad/sec$.


In all four cases the shearing wall is rotating with 10 rad/sec. This
sets the linear shearing velocity range between 80 $\av{d}$/sec (close
to inner cylinder) and 120 $\av{d}$/sec (close to outer cylinder). The
effect of different shearing velocities is discussed in
Sec. \ref{sec:nosh} and also later in Sec \ref{sec:bag}.

First we consider results without oscillations. Figure~\ref{fig:gb}(a)
shows that in the case of glued particles on the top wall, the system
reaches a state characterized by a significant shear throughout the
domain. By comparing \ref{fig:gb}(a) and \ref{fig:gb}(b), we see that
there is clearly a strong effect of glued particles on increasing
shear throughout the domain. We emphasize that the top wall without
glued particles, Fig.~\ref{fig:gb}(b), does not lead to significant
shear, although the wall itself is very frictional and inelastic, see
Table~\ref{tab:parameters}.

Figure~\ref{fig:gb}(c-d) shows the results with oscillations. The
addition of oscillations accounts for a considerable slip velocity at
the bottom wall. Similarly to the case without oscillations, the
presence of glued particles greatly reduces the slip velocity at the
top wall. Oscillations are also accountable for size segregation: The
first two bins in cases (c) and (d) are populated mostly by small
diameter particles, as it can be seen in the bottom panel. We also
note (Fig.~\ref{fig:gb}, top panel) a very small volume fraction close
to the bottom wall. This is due to the fact that when the bottom wall
moves downward, the adjacent granular layer does not follow it, except
possibly small diameter particles. This observation is also confirmed
by computer visualization of the granular layer.

The distribution of volume fraction as a function of $y$ is shown on
the same plots as velocity distribution. The steady states, marked by
filled circles, show that the volume fraction is not uniform. For each
of four cases we observe the maximum local volume fraction $\approx
58\%$ for some intermediate $y$'s, with dilation effects close to the
walls, in particular the top one.

From energy plots we estimate the equilibrating time for all four
cases: $t_{eq}(a) = 15 \, sec$, $t_{eq}(b) = 15 \, sec$, $t_{eq}(c) =
2\, sec$, $t_{eq}(d) = 6 \, sec$. These times are shorter for the
systems with oscillating bottom wall, due to increased collision
rate. Also, oscillations lead to modulation of the scaled energy; this
is shown in the insets in the energy plots. In addition, in the
systems with oscillations, the glued particles provide an increase in
collision rate and decrease in equilibration time; their effect on
equilibrating time is weak in the system without oscillations.


\subsection{Wide configurations with inelastic and frictional sidewalls}
\begin{figure*}[!ht]
\fourtitles
\hbox{
\includegraphics{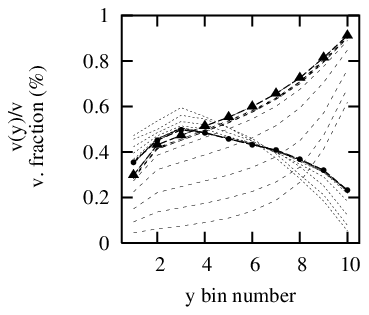} 
\includegraphics{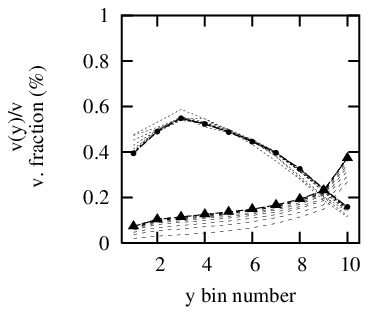} 
\includegraphics{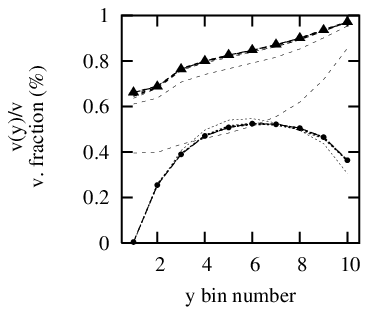} 
\includegraphics{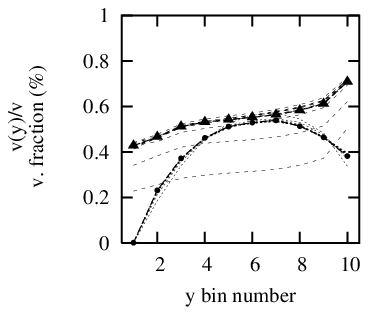} 
}
\hbox{
\includegraphics{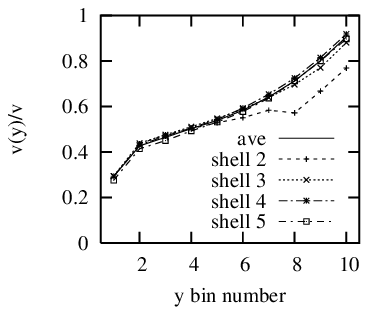} 
\includegraphics{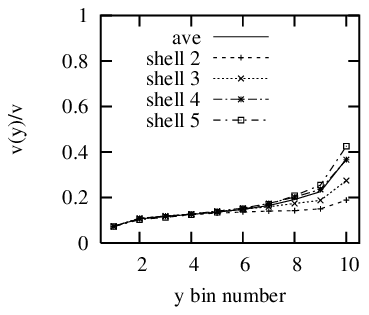} 
\includegraphics{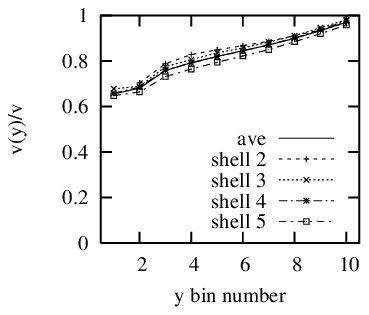} 
\includegraphics{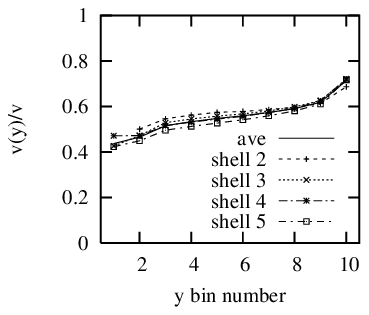} 
}
\hbox{
\includegraphics{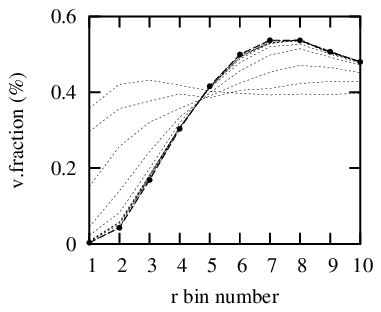} 
\includegraphics{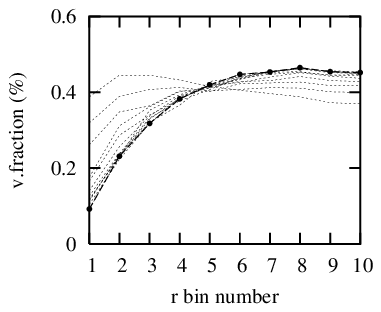} 
\includegraphics{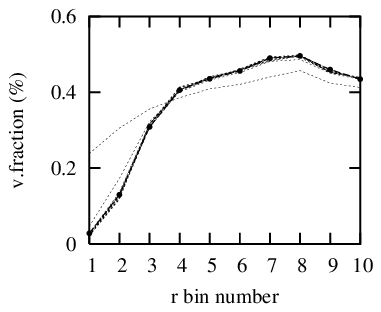} 
\includegraphics{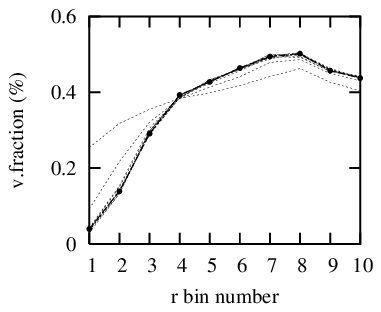} 
}
\hbox{
\includegraphics{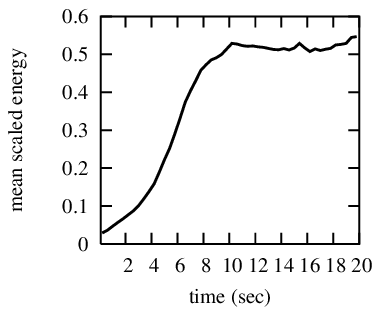} 
\includegraphics{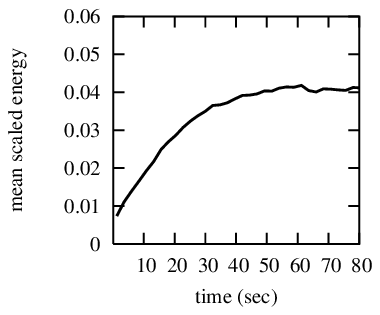} 
\includegraphics{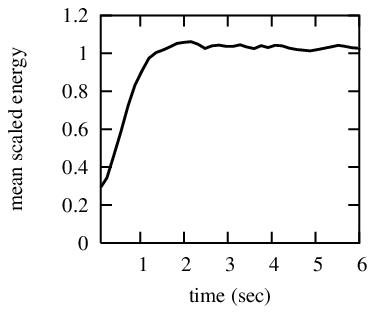} 
\includegraphics{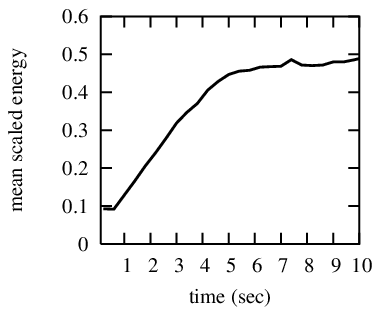} 
}
\abcd
\caption{\label{fig:wide_01} Results for wide configurations. First
Panel: scaled velocity profiles (filled triangles) and volume fraction
profiles (filled circles) in steady state regime together with the
profiles in transient regimes (dashed lines). Second panel: velocity
profiles measured in cylindrical shells (as discussed in the text)
scaled by shell's shearing velocity. Third panel: radial volume
fraction profiles. Forth panel: scaled energy plots. As in
Fig.~\ref{fig:gb}, (a) and (b) are without oscillations, (c) and (d)
are with oscillations, (a) and (c) are with glued particles on a top
wall, and (b) and (d) are without glued particles.}
\end{figure*}
Next, we consider the velocity and volume fraction profiles for wide
configuration, characterized by 5000 particles, and 40\% solid volume
fraction, using typical values of all other parameters. The results
for the same four boundary conditions as above are shown in
Fig.~\ref{fig:wide_01}.

The top panel shows the velocity and volume fraction
profiles. Qualitatively these results are similar to the results for
typical configuration shown in Fig.~\ref{fig:gb} (top~panel), although
there are some differences. For example, we see that the velocities
are larger for wide configurations. This can be explained by realizing
that the ratio of the areas of dissipative sidewalls to the area of
shearing wall is smaller. Therefore, the influence of side walls is
weaker in wide configuration compared to typical configuration. This
effect also leads to larger slip velocity at the bottom wall. On the
other hand, the volume fraction profiles are very similar in these two
configurations.

The second panel of Fig.~\ref{fig:wide_01} shows the velocity profiles
averaged over cylindrical shells of increased radius [see
Fig.~\ref{fig:ybin}(b)]. To obtain these results we split the volume
of the cell into five concentric cylindrical shells of same
thickness. Each shell is given a number $1$ to $5$ with higher number
corresponding to the larger radius: The profiles for shell $1$ are not
shown since the reduced volume fraction near inner cylinder (see the
third panel of Fig.~\ref{fig:wide_01}) does not allow for good
statistics. For comparison, the results for the velocity averaged over
the whole volume are also replotted as solid lines. These results
indicate that, as we change the location of the averaging shell inside
the cell, the changes of the velocity profiles shape are small, if not
negligible. We note here that these results are obtained assuming side
walls are characterized by relatively small friction and
dissipation. Therefore, there is no observable slow-down of the
particles in the vicinity of the side walls.

Figure~\ref{fig:wide_01}, third panel, shows the radial volume
fraction distribution obtained from averaging the volume fraction in
each cylindrical shell. As in the case of volume fraction profiles in
the vertical direction, the dashed lines mark the transient states and
the solid lines with filled circles mark the steady state
distributions. We observe significant dilation near the inner
wall. This is due to centrifugal effect: Particles colliding with the
top wall gain the momentum that, on average, has a direction along the
tangent line to the trajectory at the contact point. Following this
direction, particles get closer to the outer wall. When shearing is
strong, as in the cases (a), (c), and (d), we also observe dilation
band close to this wall.

Finally, Fig.~\ref{fig:wide_01}, forth panel, shows the scaled
energies obtained similarly to the results in
Fig.~\ref{fig:gb}~(third~panel), but averaged over longer time periods
to reduce noise (therefore, called ``mean'').  Comparing energy plots
in Figs.~\ref{fig:wide_01} and \ref{fig:gb} we see similar estimates
of the equilibrating times for all the boundary conditions, except the
one without oscillations and without glued particles. Here we notice
huge increase of equilibrating time. An explanation for this increase
is that the time period to establish steady radial volume fraction
distribution is very long, as illustrated in Fig~\ref{fig:wide_01}(b),
third panel.

%% file: base_noshaking.tex
\subsection{\label{sec:nosh} Effect of dissipation parameters and
shearing velocity in the systems without oscillations}
Next, we discuss system's behavior when we modify the dissipation or
the driving parameters. As a basis, we use the 40\% volume fraction
typical configuration system without oscillations and without glued
particles, Fig.~\ref{fig:gb}(b), with one modification: to enhance the
shearing throughout domain, we replace the dissipative side walls by
completely elastic side walls.

\begin{figure*}[!ht]
\centerline{\hbox{
\includegraphics{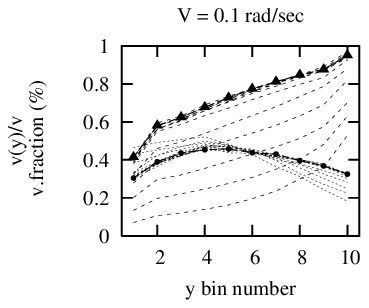}
\includegraphics{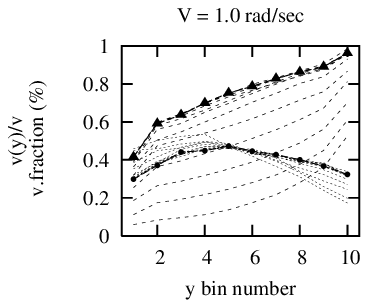}
\includegraphics{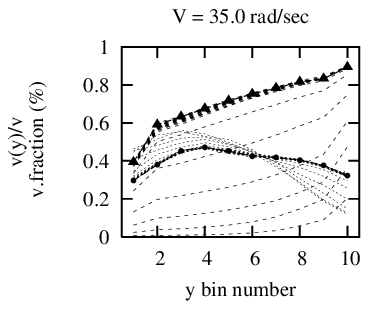}
}}
\centerline{\hbox{
\includegraphics{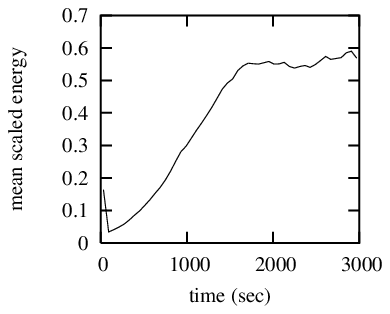}
\includegraphics{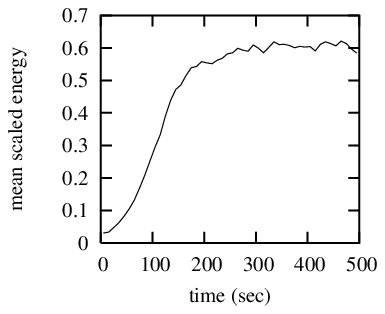}
\includegraphics{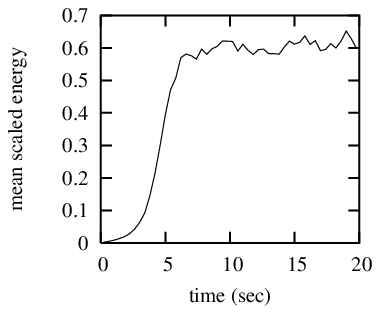}
}}
\centerline{\hbox{\hspace{0.4in} (a) \hspace{1.3in} (b) \hspace{1.4in} (c)}}
\caption{\label{fig:bnosh1} Linear-asymmetric velocity profiles for
the case of elastic sidewalls, no glued particles, and no
oscillations.  Shearing velocity is (a) $V=0.1$ rad/sec, (b) $V=1.0$
rad/sec and (c) $V=35$ rad/sec.}
\end{figure*}
\begin{figure*}[!ht]
\includegraphics{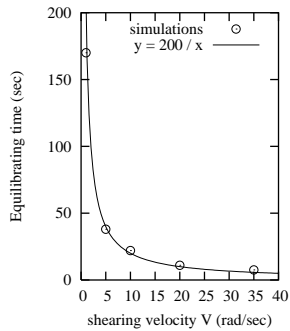}
\caption{\label{fig:bnosh3} Equilibrating times $t_e$ in simulations (open
circles) are evaluated from corresponding scaled energy plots and
plotted against shearing velocity (elastic side walls, no glued
particles, and no oscillations).}
\end{figure*}

There is a significant, qualitative difference of the velocity
profiles for the system with elastic and smooth side walls shown in
Fig.~\ref{fig:bnosh1} compared to inelastic and rough sidewalls shown
in Fig.~\ref{fig:gb}(b). Effect of shearing is much stronger in the
systems with elastic side walls. In this case, the velocity profile is
almost linear with very large slippage velocity at the bottom wall,
and almost no slippage at the top wall. We refer to this shape of
velocity profile as {\em linear-asymmetric}. The asymmetry is the
history related effect: If we shear the same initial configuration by
rotating the top wall in one direction and bottom wall in the opposite
one, but with the same magnitude of velocity $V/2$, we obtain {\em
linear-symmetric} velocity profile with equal slippage velocity at the
top and bottom.

Figure~\ref{fig:bnosh1} shows that the velocity and volume fraction
profiles are very similar as shearing velocities are
varied. Additional simulations have shown that the velocity and volume
fraction profiles are almost identical for all the tested values of
$V$ between 0.1 rad/sec and 35 rad/sec. The energy plots (bottom panel
in Fig.~\ref{fig:bnosh1}) show, however, that the time to reach the
steady state ({\em equilibrating time}), $t_e$, does depend strongly on
$V$. This effect is shown in more detail in Fig.~\ref{fig:bnosh3},
which shows that $t_e$ is inversely proportional to $V$. Therefore the
same amount of strain is needed to reach the steady state for all
explored shearing velocities.

In the next set of simulations we investigate the effect of horizontal
walls properties in the same system. Out of three parameters that
characterize the properties of horizontal walls, see
Table~\ref{tab:parameters}, we vary one at a time and keep the others
fixed. Table~\ref{tab:nosh} summarizes our findings by describing the
shearing regime for the selected ranges of the varied parameter.

\begin{table*}[!ht] 
\caption{\label{tab:nosh} Shearing regimes for different horizontal
walls properties.}
\begin{ruledtabular}
\begin{tabular}{|c|c|c|c|}
~normal restitution~ &
~tang. restitution~ &
\multicolumn{2}{c|}{~friction~}
\\ \hline
 ~$\epsilon_w$ [0.6-0.9]~& 
 ~$\beta_{0w}$ [0.0-0.9]~ &
 ~$\mu_w$ [0.0-0.33]~&
 ~$\mu_w$ [0.33-0.9]
\\ \hline
~linear-asymmetric~&
~linear-asymmetric~&
~no shearing~&
~linear-asymmetric~ \\
\end{tabular}
\end{ruledtabular}
\end{table*}

We find that $\epsilon_w$ and $\beta_{0w}$ have relatively weak effect
on the velocity profiles. However, the coefficient of friction,
$\mu_w$, influences the flow strongly. There exists a critical value
$\mu_w^c$, below which the frictional force from the shearing wall
cannot excite the granular flow. For our configuration this value is
$\mu_w^c < \approx 0.4$, for the tested range of $V$'s. In what
follows, we explain the influence of $\mu_w$ on these profiles in some
more detail.

\begin{figure*}[!ht]
\centerline{
\hbox{
\includegraphics{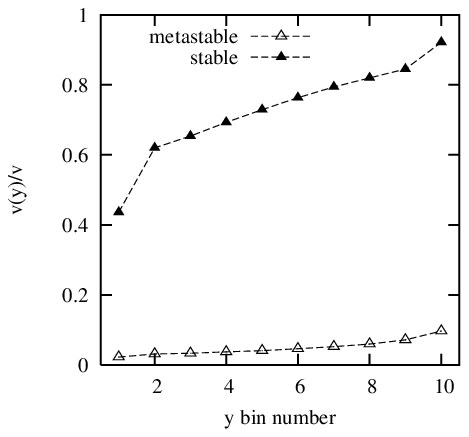}
\includegraphics{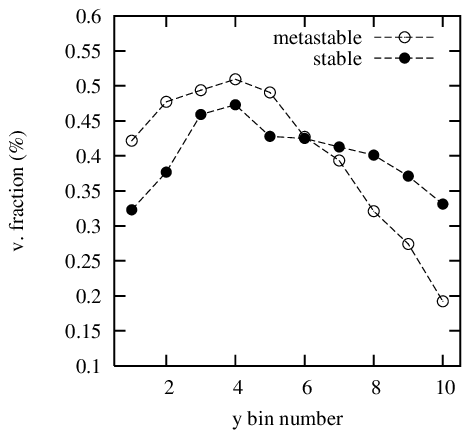}
}}
\vspace{-0.2cm}
\hbox{\hspace{2.0in} (a) \hspace{2.0in} (b)} 
\vspace{0.5cm}
\centerline{
\hbox{
\includegraphics{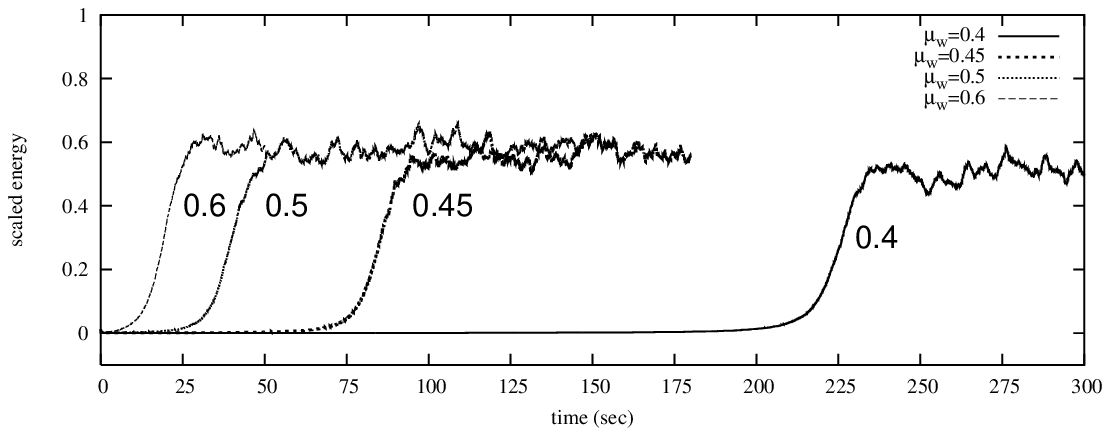}
\includegraphics{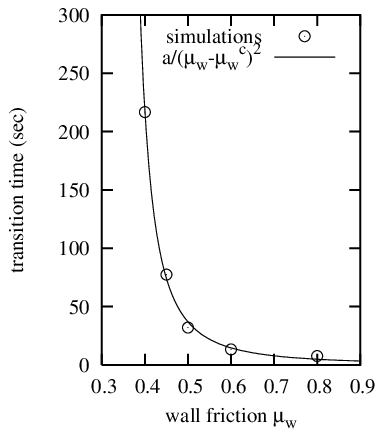}
}}
\hbox{\hspace{2.5in} (c) \hspace{2.8in} (d)} 
\caption{\label{fig:bnosh4} Delayed dynamics regime. The shearing
velocity in all cases is $V=10$ rad/sec.  (a) and (b) show the typical
velocity and volume fraction profiles in metastable and stable states,
using $\mu_w=0.45$ at $t=30 \, sec$ (metastable state) and $t=150 \,
sec$ (stable state). (c) Scaled energy as a function of time for four
values of $\mu_w$. (d) Transition times, $t_c$, versus friction: open
circles are the simulation results, and the solid line is the best fit
using inverse function $t=a/(\mu-\mu_c)^2$ with fitting parameters
$a=1.07$ and $\mu_w^c=0.33$.}
\end{figure*}

Figure~\ref{fig:bnosh4} shows the results for $\mu_w$ in the interval
$[0.4,0.9]$. Here, we observe that the systems first reaches a
metastable state of slow shearing [open symbols in
Fig.~\ref{fig:bnosh4} (a) and (b)]. This metastable state is
characterized by negligibly small slippage velocity at the bottom wall
and very large slippage at the top, as well as by high volume fraction
close to the bottom wall. As time progresses, the shear of the
granular particles is increasing very slowly. However, after certain
time the system jumps into a stable state of fast shearing with
asymmetric-linear profile, also shown in Fig.~\ref{fig:bnosh4} (a) and
(b). We call this behavior {\em delayed
dynamics}. Figure~\ref{fig:bnosh4}(c) shows the scaled energy plots
for four values of friction which illustrate this jump from metastable
to stable state.
\begin{figure*}[!ht]
\centerline{\hbox{
\includegraphics{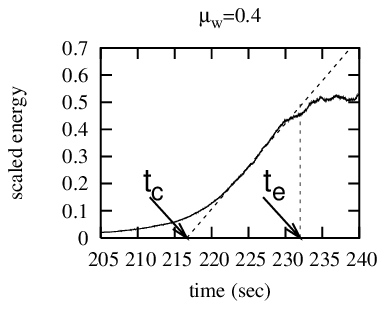}
\includegraphics{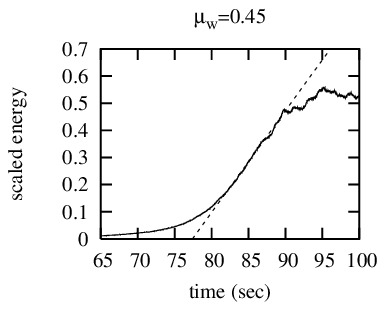}
\includegraphics{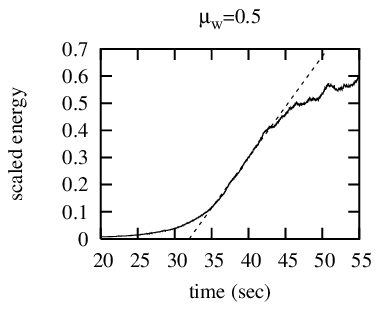}
\includegraphics{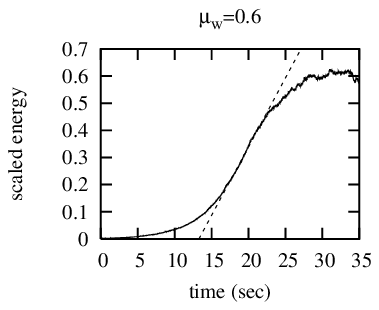}
}}
\caption{\label{fig:bnosh5} Close-up of ``transition interval'' allows to
determine the {\em transition time} $t_c$ and equilibrating time
$t_e$, as discussed in the text.}
\end{figure*}

Figure~\ref{fig:bnosh4}(d) shows the crossover {\em transition time},
$t_c$. This transition time $t_c$ (obtained in the manner discussed
below) is defined as the time when metastable state ends, {\em i.e.}
when the granular layer starts to pick up the energy fast. We find
that the inverse square function $t_c=a/(\mu_w-\mu_w^c)^2$ is a good
representation of the dependence of $t_c$ on $\mu_w$; this fit is also
shown in Fig.~\ref{fig:bnosh4}(d). The critical friction for the onset
of shearing $\mu_w^c$ is now the fitting parameter of the
function. Using a fitting routine, we obtain $\mu_w^c=0.33$.

\begin{figure*}[!ht]
\centerline{\hbox{
\includegraphics{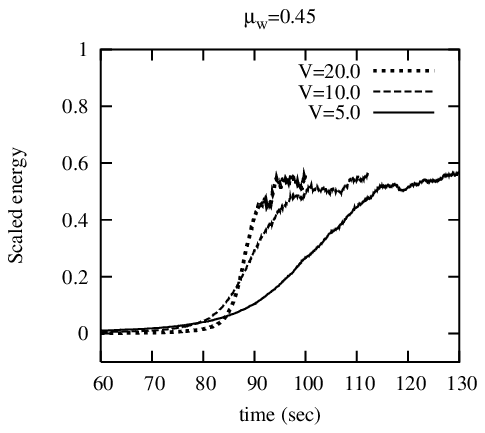}
\includegraphics{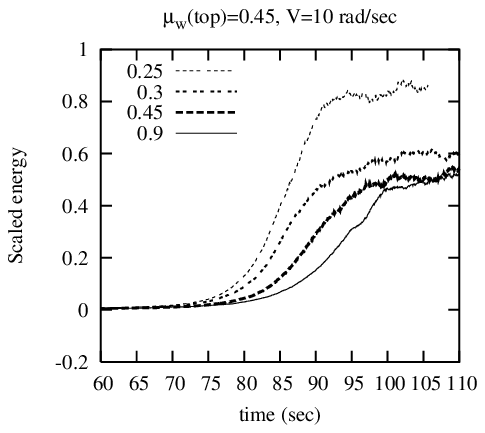}
}}
\centerline{\hspace{0.2in} (a) \hspace{1.8in} (b)}
\caption{\label{fig:mu0.45} Scaled energies in the transition
interval for (a) $\mu_w=0.45$ with three different shearing
velocities, (b) $\mu_w=0.45$ fixed for the top wall, variable $\mu_w$
for the bottom wall, fixed shearing velocity $V=10 \, rad/sec$.}
\end{figure*}

Figure \ref{fig:bnosh5} illustrates the manner in which $t_c$ is
obtained. This figure shows scaled energies with the tangent lines
(shown as skewed broken lines) at the points where the energy increase
rate is maximum. The intersection of this tangent line and the time
axis gives an estimate of $t_c$, and the intersection of the tangent
line and the horizontal line at the level of the average scaled energy
in the steady state gives an estimate of equilibrating time, $t_e$. In
all four cases, shown in Fig.~\ref{fig:bnosh5}, $\delta t_e=t_e-t_c$
is approximately $(18\pm 4)$ sec. This result is consistent with the
result for equilibrating time for the same $V=10 \, rad/sec$ and for
high friction horizontal walls ($\mu_w=0.9$), shown in
Fig.~\ref{fig:bnosh3}. For such a large $\mu_w$, $t_c \approx 0$, and
$\delta t_e \approx t_e$.

Next, we consider influence of shearing velocity for fixed friction
coefficient $\mu_w=0.45$. Figure~\ref{fig:mu0.45}(a) shows the scaled
energy in the transition time interval for the system in delayed
dynamics regime for three different shearing velocities, $V=5 \,
rad/sec$, $V=10 \, rad/sec$, and $V=20 \, rad/sec$.  This figure
confirms that the transition times $t_c$ depend very weakly on
$V$. They are all slightly scattered around the value $t_c
\approx 80 \, sec$. However, the $\delta t_e$'s are following
the inverse rule shown in Fig.~\ref{fig:bnosh3} (note that $t_c=0$ for
$\mu_w=0.9$ used in Fig.~\ref{fig:bnosh3}). Therefore, $\delta t_e$'s
strongly depend on the shearing velocity, while being almost
independent of the coefficient of friction.

To understand precisely the role which the horizontal walls play in
the mechanism of delayed dynamics, we have performed simulations with
fixed coefficient of friction of the top wall $\mu_w(top)=0.45$ and
varied coefficient of friction of the bottom wall. The results, shown
in Fig.~\ref{fig:mu0.45}(b), clearly indicate that $\mu_w (bottom)$
does not have the determining effect on the transition time $t_c$. For
example, changing it from $0.9$ to $0.25$ leads to less then 5\%
decrease of $t_c$. However, the higher slope of energy increase for
smaller friction coefficients leads to smaller $\delta t_e$. We also
note that for these low $\mu_w (bottom)$ the energies are rising to
the higher values. This is an indication of higher overall velocities
of the particles in the steady state due to higher slippage at the
bottom wall for lower $\mu_w (bottom)$.

Based on the results shown in Figs.~\ref{fig:bnosh3}-\ref{fig:mu0.45},
the mechanism of the delayed dynamics can be explained as follows.
The friction of the top wall controls the amount of the tangential
momentum given to the particles colliding with it. Thus the energy
input to the granular system is determined by the top wall friction,
as well as by the collision rate of the particles with this wall. When
the initial configuration is subjected to shear, the particles
immediately dilate close to the top wall and form the cluster of high
volume fraction close to the bottom wall. Reduced number of particles
close to the top wall then significantly decreases the input of
energy. If the friction of the top wall is low enough, below some
critical value $\mu_w^c$, all the energy is quickly dissipated without
inducing shear in the system. For higher frictions, the energy
transfered to the granular system starts to accumulate slowly: The
shearing throughout the sample increases, until, at some critical time
$t_c$, the shearing between the dense cluster and the bottom wall is
big enough to dilate a region next to this wall and to push the
cluster closer to the top wall. This process is sped up if $\mu_w
(bottom)$ is smaller. As a result, the collision rate with the top
wall increases, and the energy from this wall is rapidly transfered
into the kinetic energy of the whole cluster, until the stable state
characterized by linear asymmetric profile forms. Thus, the time span
of metastable state $t_c$ is determined mostly by the friction of the
top wall, and the time $\delta t_e$ is determined mostly by the
shearing velocity.

%% file: base_shaking.tex
\subsection{Effects of oscillations}
\begin{figure*}[!ht]
\centerline{\hbox{
\includegraphics{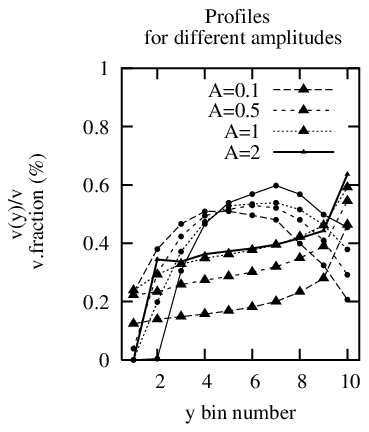}
\includegraphics{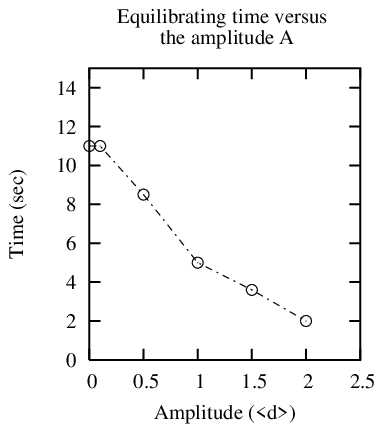}
\includegraphics{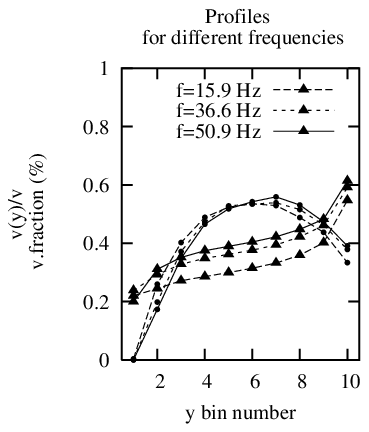}
\includegraphics{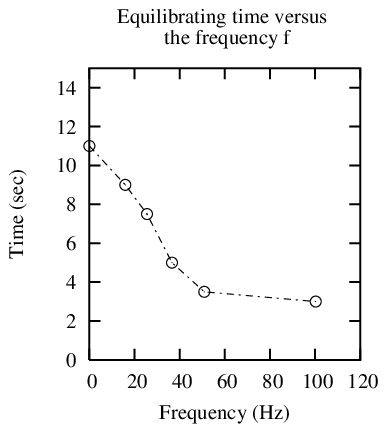}
}}
\hbox{\hspace{0.8in} (a) \hspace{1.4in} (b) \hspace{1.4in} (c)
\hspace{1.4in} (d) } 
\caption{\label{fig:bsh} 
Velocity and volume fraction profiles for (a) fixed frequency,
$\omega=\omega_t$ ($f = 36.6 \, Hz$), and different amplitudes of
oscillations; (c) fixed amplitude, $A=\av{d}$, and different
frequencies.  Also, equilibrating times are shown as a function of (b)
amplitude and (d) frequency of oscillations. The volume fraction
symbols follow the line pattern of the amplitude (a) or frequency (c)
ones.}
\end{figure*}
To test the effect of the amplitude $A$ and frequency $f$ of
oscillations, it is more convenient to go back to typical dissipative
sidewalls used in Sec.~\ref{sec:gb}. The reason is that in the case of
smooth side walls the slippage velocity at the bottom wall is
approaching the value of shearing velocity (this is shown later in
this Section), making the analysis of the velocity profiles
difficult. So, as a base system we use the ``oscillating bottom wall,
no glued particles'' configuration shown in Fig.~\ref{fig:gb}(d).

Figure~\ref{fig:bsh} shows the results for velocity, volume fraction
and equilibrating times for the systems where: (a, b) $f$ is fixed and
$A$ is varied, (c, d) $A$ is fixed and $f$ is varied. The
characteristic feature of the systems shown in Fig.~\ref{fig:bsh} is
the high volume fraction ($\nu \approx 50-60 \, \%$) band of the
granular layer in the middle of the cell height, confined between two
thin layers of low $\nu$ close to the bottom and top walls. Inside
this band the local shear rate is rather low, compared to the overall
shear rate $V/H$. The highest local shear rates are observed only
close to the top wall, while the high slippage velocities are observed
at both top and bottom wall.

Figure~\ref{fig:bsh}(a) shows the amplitude dependence of the velocity
and volume fraction profiles for fixed frequency $f=36.6 \, Hz$. We
observe that the slippage of the bottom wall depends strongly on $A$,
with higher $A$'s leading to more slippage. Also, as $A$ increases,
the peak in the volume fraction profile moves further away from the
oscillating wall. The equilibrating times, Fig.~\ref{fig:bsh}(b),
decrease with an increase of $A$ in a roughly linear manner.

Figure~\ref{fig:bsh}(c) shows the frequency dependence of the velocity
and volume fraction profiles. Here, $A$ is fixed to $1 \, \av{d}$. The
results show, again, that an increase of the intensity of
oscillations, {\em i.e.}  frequency, leads to an increase of velocity
throughout the system. However, the volume fraction profiles change
very little with $f$. The equilibrating times, Fig.~\ref{fig:bsh}(d),
decrease with an increase of $f$, with the lower rate of decrease for
higher $f$'s.

\begin{figure*}[!ht]
\hbox{\scriptsize
\hspace{0.8in} Sidewalls:
\hspace{1.0in} Sidewalls:
\hspace{0.9in} Sidewalls:  
\hspace{1.0in} Sidewalls:
}
\hbox{\scriptsize
\hspace{0.8in} typical
\hspace{1.0in} elastic-smooth
\hspace{0.8in} very frictional
\hspace{0.8in} reduced elasticity
}
\hbox{
\includegraphics{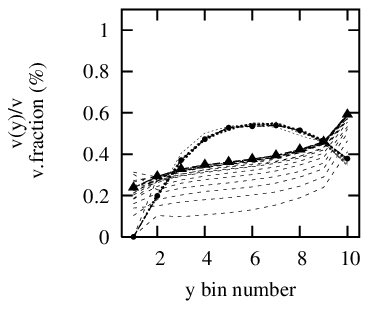}
\includegraphics{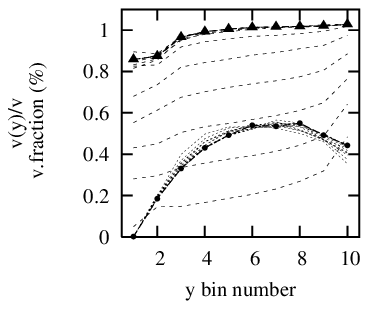}
\includegraphics{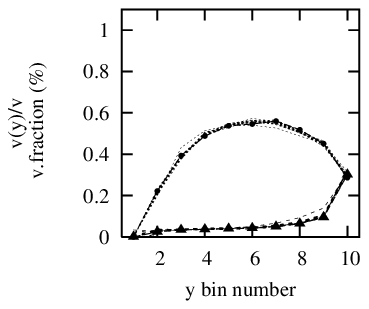}
\includegraphics{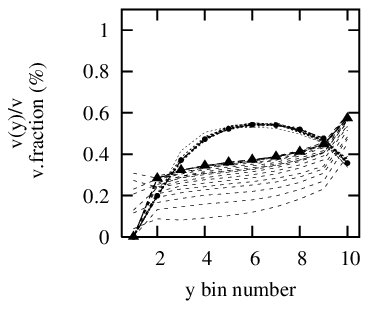}
}
\hbox{
\includegraphics{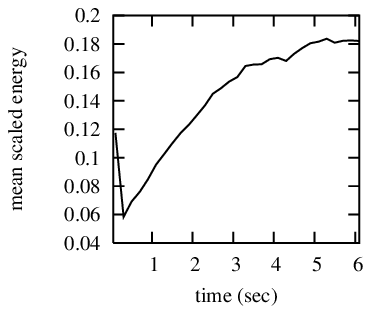}
\includegraphics{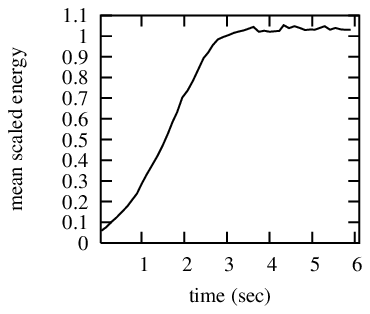}
\includegraphics{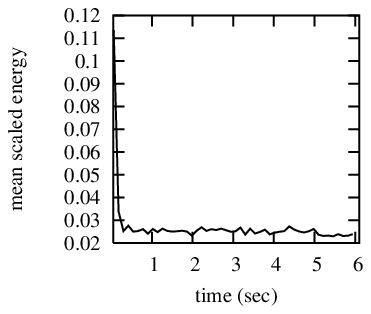}
\includegraphics{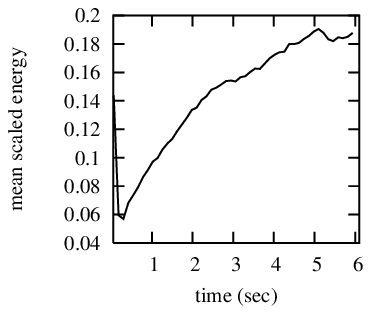}
}
\abcd
\caption{\label{fig:bsw} Effect of sidewall properties on velocity and
  volume fraction profiles. (a) Base system, $\epsilon_s=0.9$,
  $\mu_s=0.1$, $\beta_{0s}=0.35$. (b) Completely elastic and
  smooth. (c) $\epsilon_s=0.6$, $\mu_s=0.5$, $\beta_{0s}=0.35$. (d)
  $\epsilon_s=0.6$, $\mu_s=0.1$, $\beta_{0s}=0.35$. }
\end{figure*}

Previously, by comparing Fig.~\ref{fig:gb}(b) and
Fig.~\ref{fig:bnosh1}, we saw significant effect of sidewall
properties on velocity and density profiles for the systems without
oscillations. Figure~\ref{fig:bsw} shows the corresponding results as
properties of the sidewalls are modified in the case of oscillating
bottom wall, and absence of glued particles. To simplify the
comparison, Fig.~\ref{fig:bsw}(a) shows again the result for the
typical case, Fig.~\ref{fig:gb}(d).
Setting $\mu_s=0$, $\epsilon_s=1$ [Fig.~\ref{fig:bsw}(b)] makes almost
all particles move with shearing velocity. Setting $\mu_s=0.5$,
$\epsilon_s=0.6$ [Fig.~\ref{fig:bsw}(c)] creates already enough
dissipation of energy to almost completely resist the shearing force:
even oscillations cannot improve shearing when sidewalls are very
frictional. Next we explore whether it is $\mu_s$ or $\epsilon_s$ that
influence the velocity profiles. Therefore, in Fig.~\ref{fig:bsw}(d)
we show profiles and scaled energy data for $\mu_s=0.1$,
$\epsilon_s=0.6$. By comparing (a) and (d) we see that the elasticity
of the side walls does not have the observable effect on the velocity
profiles. Therefore, $\mu_s$ plays the crucial role, similarly as
observed regarding $\mu_w$ in Sec.~\ref{sec:nosh}.  The volume
fraction profiles and equilibrating times, on the other hand, very
weakly depend on the side wall properties.

%% file: base_forces.tex
\section{\label{sec:forces} Normal and shear stresses on the boundaries}
Stresses in granular systems have been the focus of many studies,
because of their importance in a number of engineering designs. A
significant part of these studies\cite{savage81,jenkins83,savage98}
deals with theoretical continuum models where stress is considered as
a mean local quantity. Therefore, the validity of these models depends
on the strength of stress fluctuations on the scale which defines
``locality''. The knowledge of stress distributions becomes crucial
here. There are many experimental, numerical, and theoretical studies
of stress
fluctuations,\cite{miller96,thornton97,radjai96b,coppersmith96}
however, these studies deal with high volume fraction of static or
slowly sheared granular systems. In these dense systems the
distribution of stresses is characterized by an exponential decay for
large stresses. It is speculated\cite{radjai96b} that the exponential
tails in stress distributions are related to the presence of force
chains in these systems. However, recently, Longhi, Easwar and
Menon\cite{longhi02} reported the experimental evidence of exponential
tails in stress distributions in rapidly flowing granular medium, {\em
i.e.} in the system where it was unlikely to find force chains in the
traditional ``static'' sense.

The results presented in this Section deal with the stress
distributions at the physical boundaries of rapidly sheared granular
systems that are not dense, {\em i.e.} 40\% volume fraction. Before
presenting these results we should note that in order to make maximum
connection with existing experimental results and theories, we
calculate the stresses in a manner that is very similar to
experimental methods, {\i.e.} we also introduce in our simulations 
stress sensors on the boundaries (more details provided below). In the low
volume fraction and/or rapid flow regime these sensors can report zero
stress during some measuring intervals, see, for example,
Ref.~\onlinecite{longhi02}. Therefore, the total distribution of
stresses must contain the distribution of {\em non-zero stresses} and
a weighted delta function to account for {\em zero stresses}. In what
follows we make a clear distinction between zero and non-zero stresses
and extend all relevant theoretical arguments to account for the
presence of zero stresses.

\begin{figure*}[!ht]
\centerline{\hbox{
\includegraphics{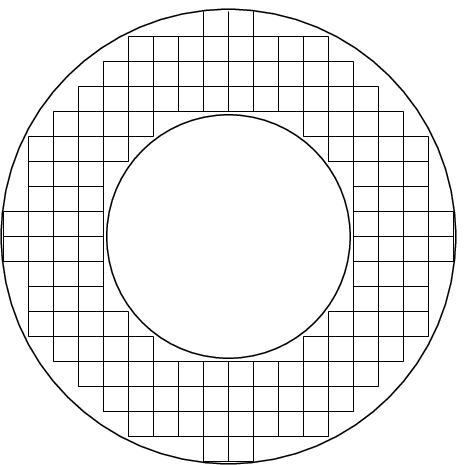}
\hspace{1cm}
\includegraphics{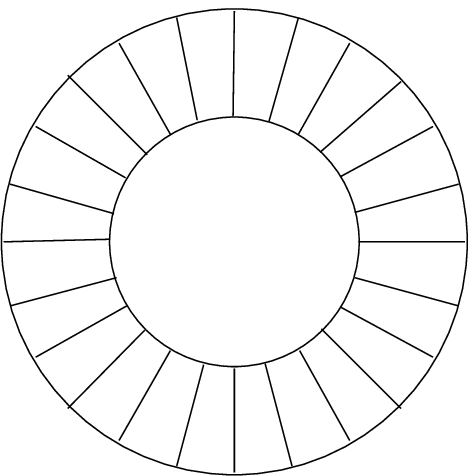}
}}
\centerline{\hbox{\hspace{0.1in} (a) \hspace{2.1in} (b) }}
\caption{\label{fig:fsensors} Stress sensor areas between inner and
  outer cylinder of Couette cell. (a) Square sensors. Only the data
  from the sensors completely inside the volume of the cell are
  used. (b) Sector-shaped sensors.}
\end{figure*}
We present the results for the stresses on the boundaries for the
typical configuration systems shown in Fig.~\ref{fig:gb}. More
detailed analysis will be given for the particular system without
oscillations and with glued particles, Fig.~\ref{fig:gb}(a). Of
particular interest here are the distributions of normal and shear
stresses on the top and bottom walls. To calculate these stresses we
cover the respective boundaries with a grid of square {\em stress
sensors}, as shown in Fig.~\ref{fig:fsensors}(a). All sensors have the
same area $A_s=d_s^2$.  Typically, we set $d_s=1.09\, \av{d}$,
$A_s=1.19\, \av{d}^2$, and analyze the data only from the sensors that
are completely inside the volume of the cell. Later in this Section we
use {\em sector shaped} sensors, Fig.~\ref{fig:fsensors}(b), which are
convenient to set large sensor areas.  The instantaneous normal and
tangential stresses on a sensor are defined in the following way:
\begin{eqnarray}
  \sigma_n = \frac{\sum_i \Delta p_n(i)}{A_s \Delta t} = 
  \frac{1}{A_s \Delta t} \frac{1}{6} \pi \rho_s \sum_i d_i^3
  |v_y^{\prime}-v_y| \label{eq:fce_n} \\
  \sigma_t = \frac{\sum_i \Delta p_t(i)}{A_s \Delta t} = 
  \frac{1}{A_s \Delta t} \frac{1}{6} \pi \rho_s \sum_i d_i^3
  (v_t^{\prime}-v_t) \label{eq:fce_t} 
\end{eqnarray}
Here $i$ counts all the collisions between the particles and the walls
in the sensor area during a time interval $\Delta t$. We refer to this
time interval as {\em averaging time}. $\Delta p_n(i)$ and $\Delta
p_t(i)$ are the change of particle's normal and tangential components
of momentum, $\rho_s $ is the density of the solid material, $d_i$ is
the diameter of a particle participating in the collision $i$, and
vectors ${\bf v}$ and ${\bf v}^{\prime}$ are the velocities of the
colliding particle before and after a collision,
respectively. Subscripts $n$ and $t$ refer to the normal and
tangential components.
Our typical choice for the time interval is $\Delta t = T_w/10$.  In
order to get sufficient data for the stress Probability Distribution
Function (PDF), we measure instantaneous stresses on all sensors every
$\Delta t$ for total of at least $\Delta T = 1000\, \Delta t$. For the
considered geometry, Fig.~\ref{fig:fsensors}(a), we have the total of
124 sensors between the inner and outer wall.
When presenting the results for stress distributions we show the
scaled histograms (PDFs) of locally measured normal or tangential
stresses, $\sigma_n$ and $\sigma_t$, as a function of
$\sigma_n/\av{\sigma_n}$ or $\sigma_t/\av{\sigma_t}$,
respectively. The averages are defined as
\begin{eqnarray}
  \av{\sigma_{(n,t)}} =(\sum_i \Delta p_{(n,t)}(i)) / A_{base} /
  \Delta T
  \label{eq:avs} \\
  A_{base}=\pi(R_{o}^2-R_{i}^2).
  \label{eq:abase}
\end{eqnarray}
Here $i$ counts all collisions between the particles and the top or
bottom wall during the whole measuring period $\Delta T=1000\, \Delta
t$.

\subsection{Normal stress distributions}
\begin{figure*}[!ht]
\fourtitles
\hbox{
\includegraphics{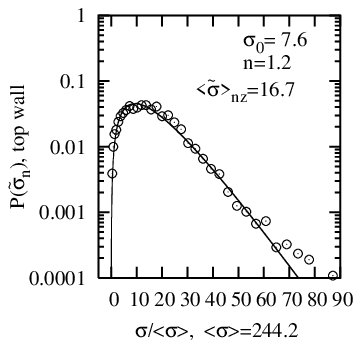}
\includegraphics{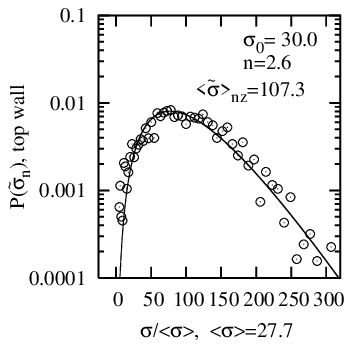}
\includegraphics{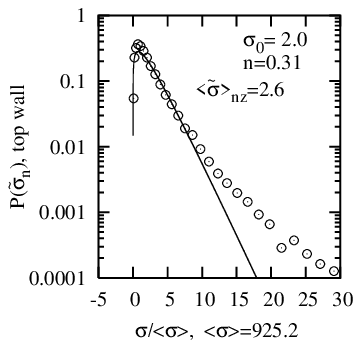}
\includegraphics{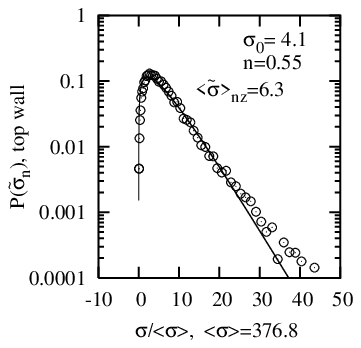}
}
\hbox{
\includegraphics{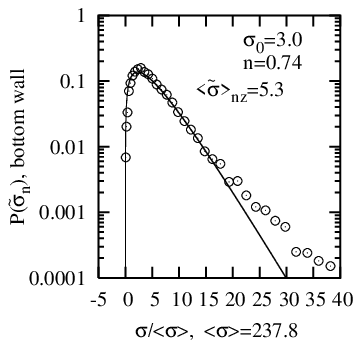}
\includegraphics{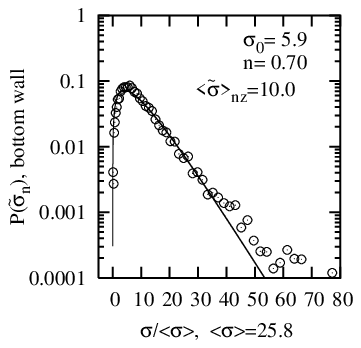}
\includegraphics{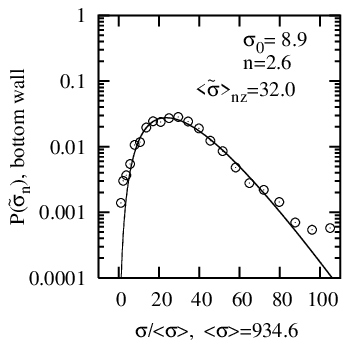}
\includegraphics{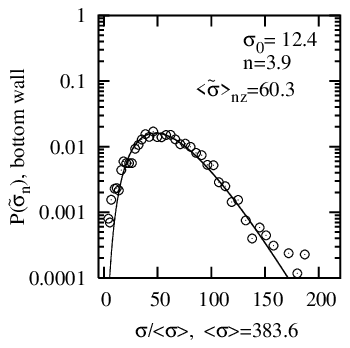}
}
\abcd
\caption{\label{fig:forces_1} Log plot of PDF for normal non-zero
stresses on the top wall (top panel) and on the bottom wall (bottom
panel) in typical configuration with four boundary conditions: (a) and
(b) without oscillating bottom wall, (c) and (d) with oscillating
bottom wall, (a) and (c) with glued particles, (b) and (d) without
glued particles. Average normal stress per area $\av{\sigma}$ is in
units of $\rho_s \av{d}^2/sec^2$. Solid lines: the best fits of the
function (\ref{eq:fdist}), with the values of the fitting parameter
$n$ and corresponding $\sigma_0$ and $\av{\sigma}_{nz}$ shown on each
plot.}
\end{figure*}
Figure~\ref{fig:forces_1} shows the results for normal {\em non-zero}
stresses on the top and bottom walls in the typical configuration for
the same four boundary conditions that were discussed in the preceding
Section (see Fig.~\ref{fig:gb}). The main features of the stress
distributions are the increase from zero for small stresses and the
decrease, in most cases approximately exponential, for large
stresses. We note that it is well established experimentally that the
exponential tails are the characteristic features of the stress PDFs
in {\em static} granular materials at volume fractions close to the
random close packing.\cite{radjai96b, thornton97, mueth98} Although
the stresses we measure are not static, nevertheless, the
approximately exponential tails are observed in all cases shown in
Fig.~\ref{fig:forces_1} (except perhaps for very large stresses). This
is evident from the best fit of the following functional form
\newcommand{\ts}{\tilde{\sigma}}
\begin{equation}
  P_{nz}(\ts)=c \ts^n e^{-\ts/\sigma_0}
  {\mbox ; \hspace{2cm}} \ts=\sigma / \av{\sigma} ,
  \label{eq:fdist}
\end{equation}
shown in Fig.~\ref{fig:forces_1} as solid lines; we use $\av{\sigma} =
\av{\sigma_n}$ as defined in (\ref{eq:avs}). The subscript ${nz}$ is
used to emphasize that the distributions are build from non-zero
stresses. This is important because certain fraction of sensors
registers zero stress due to the fact that the volume fraction
(typically 40\%) is not sufficiently large to warranty that there is
always a particle in contact with a sensor. Including zero stresses in
the definition of $P$ would prevent us from formulating the fitting
function (\ref{eq:fdist}) and relate our results directly to
experimental and theoretical results. We discuss the consequence of
this approach in more details below.


Form (\ref{eq:fdist}) can be thought of as the generalization of the
theoretical predictions for static stress distributions\footnote{The
important assumptions of theoretical models of stress distributions
are the maintained contact between particles that are closely
packed. These assumptions are valid in both completely static case and
in the case of slow shearing at high density.}. For example, the
original $q$ model proposed by Liu, Nagel, Schecter, Coppersmith,
Majumdar, Narayan, and Witten\cite{liu95,coppersmith96} and extended
later by other authors,\cite{nguyen99,socolar98} predicts the
distribution (\ref{eq:fdist}) with $n=N_c-1$, where $N_c$ is the
number of force transmitting contacts between a particle in one layer
and particles in the adjacent layer. In three dimensions and {\em fcc}
close packing $N_c=3$, so $n=2$. Using different approach, Edwards and
Grinev\cite{edwards03} predict the distribution (\ref{eq:fdist}) with
$n=1/2$. $n$ can also depend on the number of contacts between sensor
and particles during averaging time, as will be shown in
Sec.~\ref{sec:ave}. Therefore, it makes sense to think of $n$ as a
fitting parameter. Other parameters, $c$ and $\sigma_0$, can be
expressed in terms of $n$ and the average stress $\av{\ts}_{nz}$
(measured directly in simulations) using the following expressions:
\begin{eqnarray}
  1 
  &=& \int_0^\infty  P_{nz}(\ts) \, d\ts =c \, [\Gamma(n+1)\sigma_0^{(1+n)}]
  \label{eq:c1} \\
  \av{\ts}_{nz} 
  &=& \int_0^\infty \ts P_{nz}(\ts) \, d\ts = \sigma_0 (n+1)
  \label{eq:fnz}
\end{eqnarray}

In Fig.~\ref{fig:forces_1} we show the fit of the functional form
(\ref{eq:fdist}) with $n$ as the only fitting parameter. Other
parameters, calculated using (\ref{eq:c1}) and (\ref{eq:fnz}), and
average quantities, obtained directly from simulations, are also
shown. These results show that the average stresses $\av{\sigma}$
(indicated in the horizontal axes labels on each plot) at the top and
bottom walls are equal within the statistical error. However, the peak
value and the width of a distribution is different in each case. In
particular, for the systems without oscillations the distributions are
wider at the top wall, while if oscillations are present, they are
wider at the bottom wall. Before discussing this result, we first
consider the effects of volume fraction, collision rate, averaging,
and correlations on the properties of the stress distributions.

\subsubsection{The width of stress distributions}
In previous section we have shown that a distribution is determined by
the parameter $n$ and the average stresses $\av{\sigma}_{nz}$ and
$\av{\sigma}$. Here we define the width of a distribution in terms of
the above quantities and discuss its dependence on the various
properties of granular system.

Let the width of a distribution be defined as the root mean square of
all measured stresses (including zero stresses) divided by the mean
stress
\begin{equation}
\mbox{width} = \frac{\sqrt{\av{\sigma^2}-\av{\sigma}^2}}{ \av{\sigma}}
= \frac{\sqrt{\av{\ts^2}-\av{\ts}^2}}{\av{\ts}}
= \sqrt{\av{\ts^2}-1}
\label{eq:a_width}
\end{equation}
(here we use $\av{\ts}=1$). The distribution itself takes the form of
a linear combination of the distribution of non-zero stresses
(\ref{eq:fdist}) and the delta function to account for zero stresses
\begin{equation}
  P(\ts) = C_1 \, P_{nz}(\ts) + (1-C_1) \, \delta(0)
  \label{eq:true_dist}
\end{equation}
The constant $C_1$ signifies the fraction of non-zero stresses in the
total distribution. Then, the mean and the mean square of the stress
are as follows
\begin{eqnarray}
  \av{\ts}
  &=& \int_0^\infty \ts P(\ts) \, d\ts = C_1 \, \av{\ts}_{nz} = 1
  \label{eq:a_av} \\
  \av{\ts^2}
  &=& \int_0^\infty \ts^2 P(\ts) \, d\ts
= C_1 \sigma_0^2 (n+1)(n+2)
  \label{eq:a_m2} 
\end{eqnarray}
From the first equation we obtain $C_1=1/\av{\ts}_{nz}$ and, using
(\ref{eq:fnz}), we arrive at the following expression for the width of
distribution:
\begin{equation}
\mbox{width} = \sqrt{\sigma_0(n+2)-1} = \sqrt{\sigma_0+\av{\ts}_{nz}-1}
\label{eq:a_width2}
\end{equation}
Expression (\ref{eq:a_width2}) reduces to the simple
$\sqrt{\sigma_0}=(n+1)^{-1/2}$ in the case of {\em absence of zero
stresses} when all sensor register at least one strike per $\Delta t$
and $\av{\ts}_{nz}=\av{\ts}=1$. In this regime, the width is
determined by averaging effects and correlations, as we discuss
later. On the other hand, in the case {\em zero stresses are present},
we have a significant dependence of the width of a distribution on the
average non-zero stress $\av{\ts}_{nz}$. In a limiting case
$\av{\ts}_{nz} \gg 1$, appropriate to most systems shown in
Fig.~\ref{fig:forces_1}, equation (\ref{eq:a_width2}) reads:
\begin{equation}
\mbox{width} = \sqrt{ \left (\frac{1}{1+n}+1 \right ) \av{\ts}_{nz}-1}
\propto \sqrt{\av{\ts}_{nz}}
\label{eq:w2}
\end{equation}
where we have used the fact that for $\av{\ts}_{nz} \gg 1$ sensors do
not register more then one particle at a time and for single particle
contacts $n=\mbox{constant}$, as explained in next Section. Equation
(\ref{eq:w2}) is consistent with visual estimate of the widths of the
distributions in Fig.~\ref{fig:forces_1}: The larger $\av{\ts}_{nz}$
(shown in the right top area on each plot), the wider is the
distribution. Next we show that ${\av{\ts}_{nz}}$ in turn depends on
the collision rate, area of the sensors and averaging time.

Let $N_t=n_s n_T$ be the total number of stress measurements, obtained
from all $n_s$ sensors during $n_T$ time-intervals. Out of $N_t$
stresses, $N_{nz}\le N_t$ are non-zero. The non-zero average stress
depends on the size of sensor $A_s$ and averaging time $\Delta t$,
\begin{eqnarray}
\av{\ts}_{nz}=\frac{\av{\sigma}_{nz}}{\av{\sigma}} =
\frac{N_t}{N_{nz}}=\frac{1}{w \, A_s \, \Delta t }
\label{eq:avsigma}
\end{eqnarray}
where $w$ is the frequency of non-zero stress events per unit area. In
the case of low collision rate with the sensors, when no more then one
particle strikes $A_s$ during $\Delta t$, $w$ has a meaning of
particle-wall collision rate per unit area.

Relations (\ref{eq:w2}) and (\ref{eq:avsigma}) allow us to explain the
distributions in Fig.~\ref{fig:forces_1}. The sensor areas and
averaging times are fixed so it is the rate of non-zero events $w$
which determines the average and width of a distribution. To the first
approximation $w \propto$ the volume fraction close to the respective
boundary. From Fig.~\ref{fig:gb} (top panel), in the systems without
oscillations the volume fraction is smaller at the top wall compared
to the bottom wall. Therefore, the distributions are wider at the top
compared to the bottom. In the systems with oscillations the volume
fraction is smaller at the bottom wall compared to the top wall,
therefore, the distributions are wider at the bottom.

\subsubsection{\label{sec:ave} Averaging effects}
\newcommand{\tF}{\tilde{F}}

In this section we study one of the factors that can affect the
parameter $n$, {\em i.e.} the number of contacts, $N$, between the
particles and a sensor during the averaging time $\Delta t$.

Let us define the parameter $n_0$ that is characteristic for the
distribution of stresses due to single particle contacts:
\begin{equation}
  P_N(\ts) = c \ts^{n_0} e^{-\ts/\sigma_0}
  \mbox{;} \hspace{1cm} N=1
 \label{eq:n0}
\end{equation}
If more then one particle collide with the sensor of area $A_s$ during
time $\Delta t$ ($ N > 1$), then the stress distribution is different
due to {\em averaging effects}. For instance, when two particles
strike the same sensor during $\Delta t$, $N=2$, the probability of
registering total stress $F$ depends on the probability of one of them
contributing the stress $\sigma \le F$ and another one the stress
$F-\sigma$. Assuming both considered particles have independent
distributions (\ref{eq:n0}), the distribution of their collective
stress $F$ is $P_2(F)=\int_0^{F} P(\sigma)P(F-\sigma)d\sigma$. Using
(\ref{eq:n0}) and the integral identity
\begin{equation}
\int_0^F\sigma^{{r}}(F-\sigma)^{{q}}d\sigma = F^{({r}+{q}+1)} \,
\Gamma({r}+1) \Gamma({q}+1) / \Gamma({r}+{q}+2) 
\label{eq:id1}
\end{equation}
valid for any positive real numbers ${r}$ and ${q}$, and in particular
for ${r}={q}=n_0$, we arrive at the following distribution for the
stresses generated by double strikes:
\begin{equation}
P_2(\tF) = c_2 \tF^{(2n_0+1)} e^{-\tF/\sigma_0} {\mbox ; }
\, \, \tF=F / \av{\sigma} \nonumber
\end{equation}
where $c_2$ is a normalization coefficient. 
Applying the same argument to the case when $N$ particles strike the
sensor during $\Delta t$ we obtain the following distribution of
stresses:
\begin{equation}
  P_N(\ts) = c \ts^{(N(n_0+1)-1)} e^{-\ts/\sigma_0}
 \label{eq:fdist2}
\end{equation}
The above calculation sets the relation between the parameter $n$ and
the number $N$ as $n=N(n_0+1)-1$, see (\ref{eq:fdist}). Therefore,
under the assumption that particles are not correlated we obtain that
the distributions of stresses generated by multiples collisions are
characterized by higher power $n$. We will see below that this
assumption is justified for the results shown in
Fig.~\ref{fig:forces_1}.

This conclusion may be used to explain the increased values of $n$
observed in the distribution of stresses on the bottom wall in the
systems shown in Fig.~\ref{fig:forces_1}(c) and
Fig.~\ref{fig:forces_1}(d). In these systems the bottom wall is
vibrated and most of the stress data are collected during the phase
when the bottom is rising up. This phase is definitely characterized
by increased compaction of particles close to the sensors and an
increased number $N$.

However, an increase of $N$ cannot be responsible for the large $n$'s
observed on the top wall in the system shown in
Fig.~\ref{fig:forces_1}(b), and, to lesser extend,
Fig.~\ref{fig:forces_1}(a).  In particular regarding 
Fig.~\ref{fig:forces_1}(b), we expect that the source of large $n$ 
lies in the separately verified result that the typical normal 
components of the velocities of the particles colliding with the
top wall is large, leading to different stress distribution.  We note,
however, that the values of $n$ given in Fig.~\ref{fig:forces_1}(a-b)
are not as accurate as the rest of $n$'s shown in Fig.~\ref{fig:forces_1}
due to small volume fraction there (see Fig.~\ref{fig:gb}(a-b)). Therefore, 
the number of collisions is small -- e.g., we count approximately $500$ 
collisions during the whole time span of the simulations presented in
Fig.~\ref{fig:forces_1}(b).



The results of Fig.~\ref{fig:forces_1} allow us to estimate the
parameter $n_0$ defined earlier in this Section as the value of $n$
corresponding to the distribution of stresses registered by the
sensors with $N=1$. Because $n_0 \le n$ (see (\ref{eq:fdist2})), the
smallest found $n$ provides an upper bound on $n_0$. The lowest values
of $n$ are found in the distributions shown in
Fig.~\ref{fig:forces_1}: (a) bottom, (b) bottom, (c) top and (d)
top. For these cases $n<1$, therefore $n_0<1$. This result is very
different from $n_0=2$ in the $q$-model: this should be no surprise
because $q$-model assumes high volume fraction and static
configuration.

\subsubsection{\label{sec:corr} Correlations}

The dependence of the normal stress distributions on the number of
contacts per sensor can be described by (\ref{eq:fdist2}) only when
the particles participating in contact with a sensor are not
correlated. However, there are certain regimes and conditions of
granular flow when this assumption may not be correct. For example,
Miller, O'Hern and Behringer\cite{miller96} studied the response of
the force distribution to the number $N$, controlled in their case by
the size of particles, and found that in dense granular flow the
distribution widths were approximately independent of $N$. This
independence was attributed to the presence of correlations due to
presence of force chains.

In what follows we present the systematic study of the effect of
sensor size (and, hence, the number $N$) on the bottom wall stress
distributions for our typical configuration without oscillations and
with glued particles, Fig.~\ref{fig:gb}(a). This study will allow us
to confirm the validity of our predictions (\ref{eq:a_width2}) and
(\ref{eq:fdist2}) in a more quantitative way and to check for possible
correlations.

In order to explore large range of sensor sizes, here we use {\em
sector-shaped} sensors, obtained by dividing the bottom wall area by a
number of radial lines drawn from the center of the cell toward
outside boundary, see Fig.~\ref{fig:fsensors}(b). Each sensor has the
same radial dimension equal to the distance between the inner and
outer cylinder, and the same angular dimension, set to the values
${2\pi}/{m}$, $m \in [2,252]$. Therefore, the area of sector-shaped
sensors varies between 1.0 and 125.6 (in units of $\av{d}^2$), with
the largest sensor being half of the bottom wall.  For selected values
of sensor areas we have confirmed that the results do not depend on
the sensor shape. We note that all the results presented so far were
obtained using small sensors of the area $1.19$ $\av{d}^2$.


\begin{figure*}[!ht]
\centerline{
\hbox{
\includegraphics{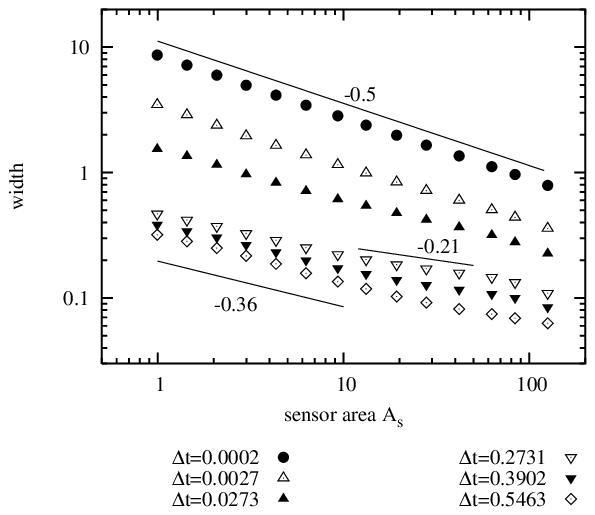}
\includegraphics{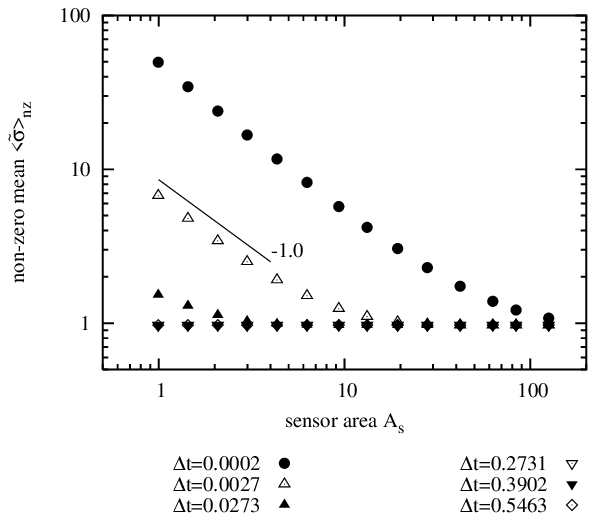}
}}
\centerline{\hbox{\hspace{0.6in} (a) \hspace{1.6in} (b) }}
\centerline{
\hbox{
\includegraphics{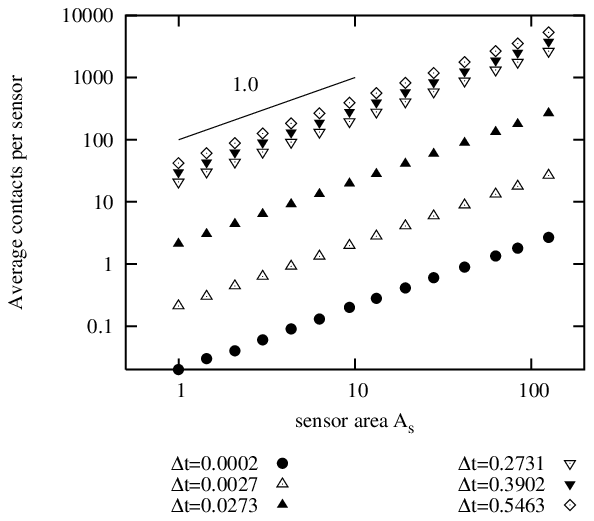}
\includegraphics{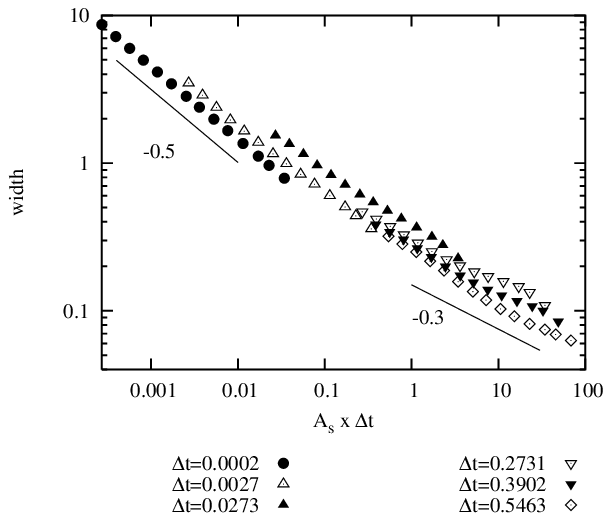}
}}
\centerline{\hbox{\hspace{0.6in} (c) \hspace{1.6in} (d) }}
\caption{\label{fig:rms1} (a) The widths of force distributions as
defined by (\ref{eq:a_width}) versus the sensor area $A_s$ (in units
of $\av{d}^2$) for six values of $\Delta t$ (shown below the
plot). (b) Average non-zero stress $\av{\ts}_{nz}$ as a function of
$A_s$ for the same six averaging times (for $\Delta t \ge 0.2731$
$\av{\ts}_{nz}=1$). (d) Average number of contacts per sensor as a
function of $A_s$. (c) Collapse of the width data against the product
$\Delta t A_s$. The solid lines have specified slopes.}
\end{figure*}
Figure~\ref{fig:rms1}(a) shows the results for width of distribution
(\ref{eq:a_width}) as a function of $A_s$ for fixed values of $\delta
t$ in the range between $T_w/100=0.0002 \, sec$ and $20 \,
T_w=0.5463$. This range includes our typical value $\Delta t =
T_w/10=0.0027$ (open triangles) used to obtain the distributions in
Fig~\ref{fig:forces_1}. To help interpretation of the results, we plot
in Fig.~\ref{fig:rms1}(b) the value of $\av{\ts}_{nz}$, in
Fig.~\ref{fig:rms1}(c) the average number of contacts per sensor, and
in Fig.~\ref{fig:rms1}(d) the widths of distributions versus the
product of the averaging time and the sensor area.

This figure shows that if zero stresses are present,
i.e. $\av{\ts}_{nz}>1$, the widths scale with the sensor area as
$A_s^{-0.5}$, which is the scaling predicted by (\ref{eq:w2}).  This
result applies to all sensor sizes when $\Delta t = 0.0002 \, sec$, to
the sensors of $A_s < 20$ when $\Delta t = 0.0027 \, sec$, and to the
sensors of $A_s < 2$ when $\Delta t = 0.0273 \, sec$, see
Fig.~\ref{fig:rms1}(b).

When zero stresses are not present ($\av{\ts}_{nz}=1$) the width
(\ref{eq:a_width2}) reduces to:
\begin{equation}
  {\mbox width} = \sqrt{\sigma_0}=(N(n_0+1))^{-0.5}
  \label{eq:width3}
\end{equation}
Here $N \propto A_s$ is a number of contacts per sensor from
(\ref{eq:fdist2}) and absence of correlations is assumed. This
relation (\ref{eq:width3}) is consistent with the $A_s^{-0.5}$ in
Fig.~\ref{fig:rms1}(a) in the case $\Delta t = 0.0027 \, sec$ and $A_s
> 20$. However, when $\Delta t \ge 0.0273$ (for these large $\Delta
t$'s, the zero stresses are absent for almost all considered sensors,
see Fig.~\ref{fig:rms1}(b)) the widths decrease with an increase of
$A_s$ at a slower rate then predicted by (\ref{eq:width3}). We observe
the slowest decrease, approximately $A_s^{-0.21}$, for the sensor
sizes $10 < A_s < 100$, and for the averaging times of $0.2731 \, sec
$.

We expect that the origin of this slower decrease of the widths is the
presence of correlations between the particles striking the same
sensor (which may include self-correlation, i.e., the same particle
colliding with a same sensor on multiple occasions during a given
$\Delta t$).  For example, the distribution of stresses due to N
completely correlated particles acting on each sensor would have
$n=n_0$, leading to the widths independent of $A_s$. Therefore, the
stronger are the correlations between particles, the lower is the
exponent in the width versus $A_s$ function. In Fig.~\ref{fig:rms1}(a)
we can see that stronger correlation effects ({\em i.e.} slower
decrease of the widths with an increase of $A_s$) occur for longer
$\Delta t$ and larger $A_s$. This is even more evident when the width
is plotted against the product $\Delta t$ times $A_s$, see
Fig.~\ref{fig:rms1}(d): the slope changes gradually from $-0.5$ for
small $\Delta t \, A_s$ to about $-0.3$ for large $\Delta t \,
A_s$. The collapse of almost all data points on approximately single
curve in Fig.~\ref{fig:rms1}(d) indicates the fact that the same
averaging or correlation effects result if either $\Delta t$ or $A_s$
is increased by the same factor.  Possible exception to this rule are
very long $\Delta t$'s, which we discuss below.

We note that Miller, O'Hern and Behringer,\cite{miller96} have found
evidences of strong correlations when they estimated experimentally
the $width \propto A_s^0$ for $\Delta t =0.0005 \,sec$ and $5 \le A_s
\le 80$. However the overall volume fraction of their samples was
considerably higher (because of the heavy top wall supported in
gravitational field by the granular material), therefore the spatial
correlations can be explained if one assumes the normal stress on the
bottom wall is applied through the network of {\em force
chains}.\cite{howell99} The results presented here are obtained for
smaller volume fraction and in the absence of gravity. However, in our
case we expect that spatial correlations may arise from statistical
fluctuations of the local volume fraction. These fluctuations lead to
occasional formation of the denser structures of particles, stretched
from the top to bottom wall: the presence of friction between the
particles insures the lifetime of these formations is relatively
long. As in the case of force chains, the dense structures carry more
stress then the ``free'' particles in the ``inter-structure'' space,
thus introducing spatial correlations. Unlike the force chains, the
dense structures do not form dense force network and are not present
always.  They may form and disappear allowing for longer
``structure-free'' periods. This concept is consistent with the fact
that our correlations are weak and the correlation regimes are
detected only when $\Delta t$ and $A_s$ are large enough to include
such a structure event.  In addition the average number of contacts
with sensors, see Fig.~\ref{fig:rms1}(c), is larger then 10 in the
regimes where (\ref{eq:width3}) fails. This fact suggest the
possibility of {\em frequent returning contacts} of the same particle
with a sensor, which in event-driven simulations is a signature of a
particle pressed to the sensor from above, as one would expect for a
particle that is member of a dense structure. The existence of similar
structures (named as ``transient force chains'') was also suggested in
Ref.~\onlinecite{longhi02}.

We note that the dependence of the width on $A_s$ becomes stronger
again in the case of very long $\Delta t$'s, see the results for
$\Delta t = 0.5463$ in Fig.~\ref{fig:rms1}(a). Simple estimate shows
that this change in trend occurs when $\Delta t$ is so long that a
typical particle, moving with a typical velocity, travels across a
sensor in the time that is shorter than $\Delta t$, therefore
decreasing this self-correlating effect.  We currently investigate the
scenarios proposed here in more detail and plan to present the results
elsewhere.

\subsection{Tangential stress distributions}
\begin{figure*}[!ht]
\fourtitles
\hbox{
\includegraphics{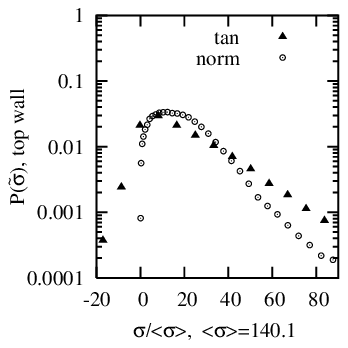}
\includegraphics{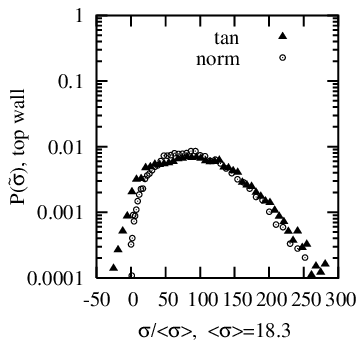}
\includegraphics{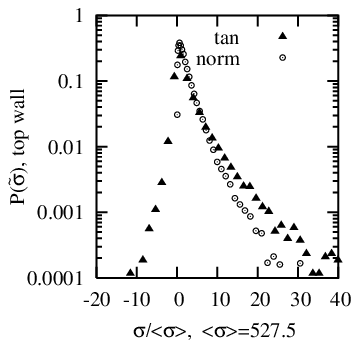}
\includegraphics{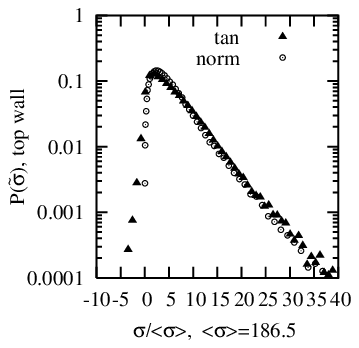}
}
\hbox{
\includegraphics{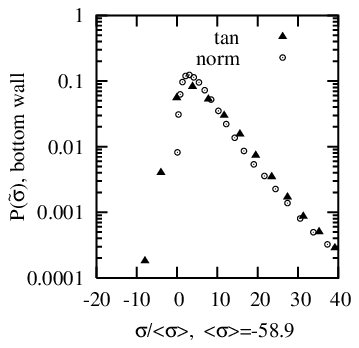}
\includegraphics{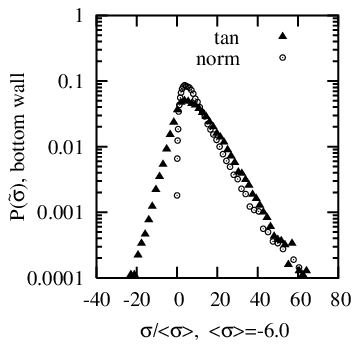}
\includegraphics{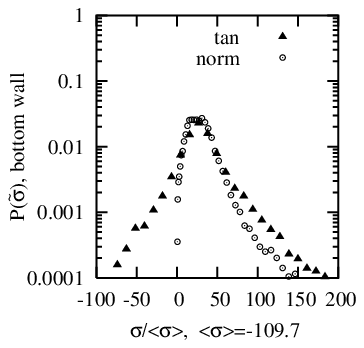}
\includegraphics{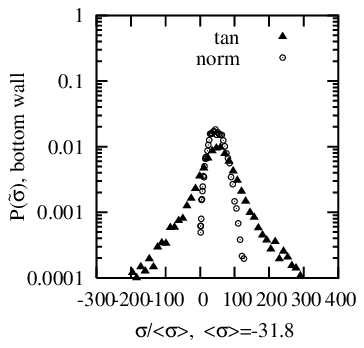}
}

\abcd
\caption{\label{fig:forces_2} Log plot of PDF for tangential non-zero
stresses (filled triangles) on the top wall (top panel) and on the
bottom wall (bottom panel) in the typical configuration with four
boundary conditions: (a) and (b) without oscillating bottom wall, (c)
and (d) with oscillating bottom wall, (a) and (c) with glued
particles, (b) and (d) without glued particles. Average tangential
stress per area $\av{\sigma}=\av{\sigma_t}$ is in units of $\rho_s
\av{d}^2/sec^2$. For comparison, normal stress distributions from
Fig.~\ref{fig:forces_1} are replotted here (open circles).}
\end{figure*}
Tangential (shear) stress is defined by
(\ref{eq:fce_t}). Figure~\ref{fig:forces_2} shows the tangential
stress distributions (filled triangles) scaled by mean tangential
stress, $\ts = \sigma_t / \av{\sigma_t}$ for the same boundary
conditions as discussed in Fig.~\ref{fig:forces_1}. For the
convenience of comparison, the normal stress distributions from
Fig.~\ref{fig:forces_1} are also shown (open circles).

The average tangential stress, shown in the horizontal label on each
plot, is positive on the top wall and negative on the bottom. This
reflects the fact that the top wall is mostly accelerating the
particles while the bottom wall is slowing them down. The absolute
value of the average tangential stress $\av{\sigma_t}$ tells us the
magnitude of torque and power needed to keep the top wall rotating and
the bottom wall stationary. It is considerably larger for the systems
with glued particles compared to the ones without glued particles,
since the glued particles enhance momentum exchange between the wall
and the free particles. We also note that $|\av{\sigma_t}|$ is larger
on the top wall then on the bottom by a factor of $~3$ for the systems
without oscillations, and by a factor of $~5$ for the systems with
oscillations. This factor is larger in the latter system since
oscillations increase the vertical velocity fluctuations and the
dissipation of the momentum on the sidewalls is increased. Since the
momentum conservation requires that the momentum applied to the system
through the top wall be equal to the momentum dissipated on both
sidewalls and on the bottom wall, less momentum is left for the bottom
wall in the systems with oscillations.

Similarly to the normal stresses, we observe the exponential tails for
$\ts > \ts_{0}$ where $\ts_{0}$ is the location of the peak of a
distribution. For most considered cases, these tails have the same
slope and location as the tails of the normal stresses. This suggests
strong correlation between the normal and tangential
stresses. However, for $\ts < \ts_{0}$ the distribution of tangential
stresses differs qualitatively from the distribution of normal
stresses. The PDF's of tangential stresses decrease to zero in an
exponential manner, allowing for stresses in the direction opposite
$\av{\sigma_t}$. However, in most cases this different behavior has
only weak influence on the width of the distributions.

An exception is the case of oscillating bottom wall
(Fig.~\ref{fig:forces_2}(c) and (d) bottom panel) where the tangential
distributions are noticeably wider then the normal stress
distribution. This can be explained by noticing that when bottom is
moving up, the particles next to the bottom wall are constrained by
increased volume fraction to move predominantly in the horizontal
directions, therefore increasing fluctuations of tangential stresses.



\begin{figure*}[!ht]
\fourtitles
\hbox{
\includegraphics{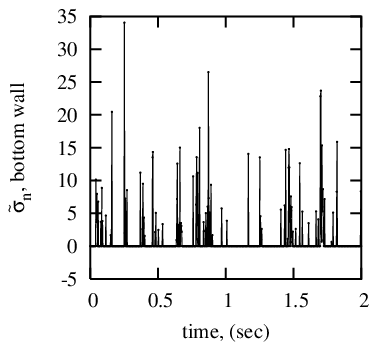}
\includegraphics{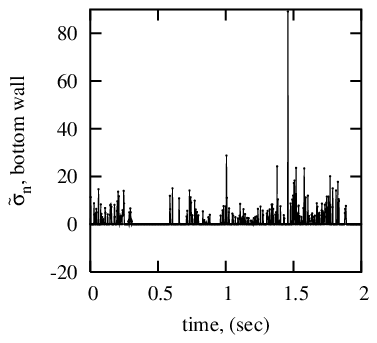}
\includegraphics{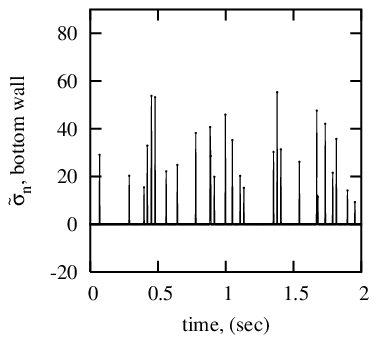}
\includegraphics{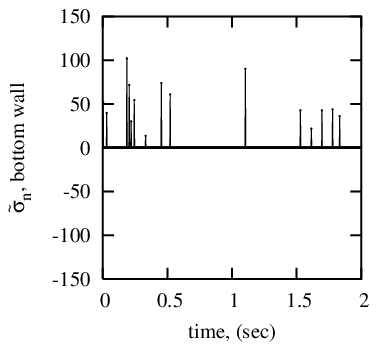}
}
\hbox{
\includegraphics{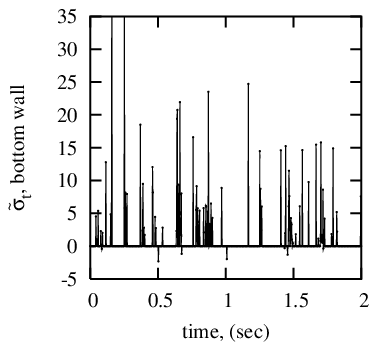}
\includegraphics{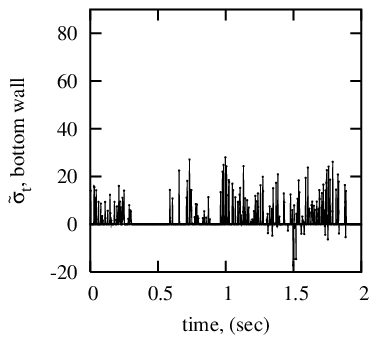}
\includegraphics{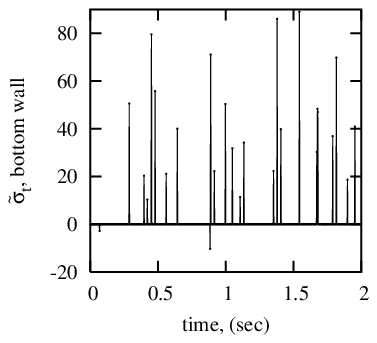}
\includegraphics{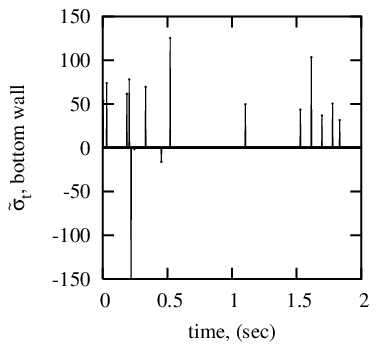}
}
\hbox{
\includegraphics{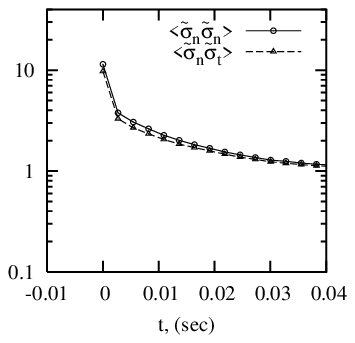}
\includegraphics{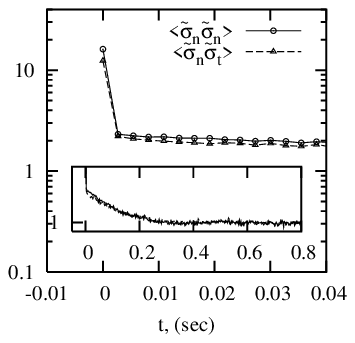}
\includegraphics{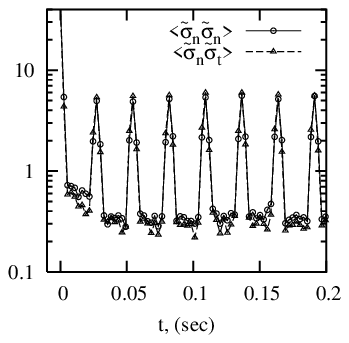}
\includegraphics{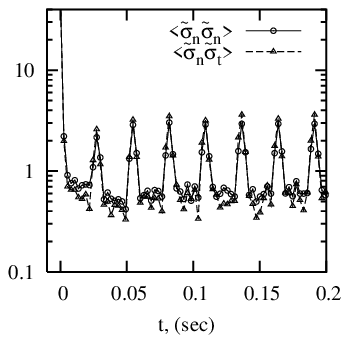}
}
\abcd
\caption{\label{fig:time_sig} First panel: normal stress signal on the
bottom wall during randomly chosen interval of 2 seconds. Second
panel: tangential stress signal on the bottom wall for the same 2
seconds interval. Third panel: stress auto- and cross-correlation
functions. The inset in (b) shows the correlation functions on the
longer time scale. (a) and (b) - without oscillating bottom wall, (c)
and (d) - with oscillating bottom wall, (a) and (c) - with glued
particles, (b) and (d) - without glued particles.}
\end{figure*}

To better understand the relation between the normal and tangential
stresses we study in more details the time series of these stresses.
Figure~\ref{fig:time_sig}, top two panels, shows the values of
$\sigma_n/\av{\sigma_n}$ and $\sigma_t/\av{\sigma_t}$ as a function of
time registered by a single sensor on the bottom wall for the four
boundary conditions as in Fig.~\ref{fig:forces_2}.
Simple visual inspection of Fig.~\ref{fig:time_sig}, top two panels,
shows that the normal and tangential stresses are correlated. This
observation is confirmed by comparing the normal stress
autocorrelation function with the appropriate cross-correlation
function, as we discuss next.

Consider the time signals $\ts_n(t)$ and $\ts_t(t)$ shown in
Fig.~\ref{fig:time_sig}, first and second panels. If these signals are
correlated then we can write
\begin{equation}
\ts_t(t) = K\,\ts_n(t) + R(t) \mbox{,} \label{eq:corr}
\end{equation}
where $R(t)$ is a noise term such that $\av{R}=0$ and
$K=\av{\ts_t}/\av{\ts_n}=1$. Then the cross-correlation function
defined as $\av{\ts_n(t_0)\ts_t(t_0+t)}$ must be equal to the
autocorrelation function defined as $\av{\ts_n(t_0)\ts_n(t_0+t)}$. In
the definition of these time correlation functions, averaging is
performed over all initial times $t_0$. To reduce noise we also
average the results of all sensors.

Figure~\ref{fig:time_sig}, third panel, shows that to within small
error $\av{\ts_n\ts_t}=\av{\ts_n\ts_n}$, therefore, confirming the
relation (\ref{eq:corr}) with $\av{R}=0$. This figure also shows that
the correlation functions decay very fast within an interval shorter
than averaging time. For longer times, the correlation function
decreases until it reaches its uncorrelated value equal to unity in
the systems without oscillations, (a) and (b), and it oscillates
around this uncorrelated value with the frequency of driving
oscillations for (c) and (d). We note that larger particle velocities
lead to faster decay of correlation functions in
Fig.~\ref{fig:time_sig}(a) compared to \ref{fig:time_sig}(b) (see also
Fig.~\ref{fig:gb}(a-b), top panel). The uncorrelated value for both
correlation functions is equal to unity because an average of product of
two uncorrelated stresses is equal to the product of average stresses.

The results of this section confirm strong correlation between the
normal and tangential stresses. This correlation renders that many
conclusions obtained for normal stresses also apply to the tangential
stresses.

\section{\label{sec:cp} Constant volume versus constant pressure simulations}
All the results we have discussed so far are obtained for the systems
characterized either by fixed constant volume, as in the simulations
without oscillations, or by prescribed volume dependence on time, as
in the simulations with oscillating bottom wall. We refer to these cases
as {\em controlled volume simulations}. A special case of controlled
volume simulations are the {\em constant volume (CV) simulations}
where the volume of the system is kept strictly constant (no
oscillations).

It is, however, well known, that the volume constraint can
significantly affect the granular flow at high volume fractions. The
system can be easily locked in a jammed state. That is why dense
sheared granular flow experiments are commonly carried out using
adjustable volume, but controlled pressure. Miller, O'Hern and
Behringer,\cite{miller96} for example, report the results for the
sheared granular flow in a Couette cell, where the weight of the top
wall provides constant pressure. Also, all free surface granular
flows in gravitational field, for example, the flow of the granular
matter down an inclined plane,\cite{louge01,silbert01} are the
examples of controlled pressure scenarios. In these cases the weight
of the granular matter itself controls the pressure both inside the
system and on the boundaries. Recently, new studies emerged that
indicated the importance of the difference between controlled volume
and controlled stress boundary conditions. For example, Aharonov and
Sparks\cite{aharonov99,aharonov02} use the 2D numerical simulations to
study the response of the dense sheared granular system to the applied
boundary pressure. They find different shearing modes determined by
pressure; These modes can have very similar volume fractions, but
different microstructural organization.

\begin{figure}[!ht]
\centerline{\hbox{ \includegraphics{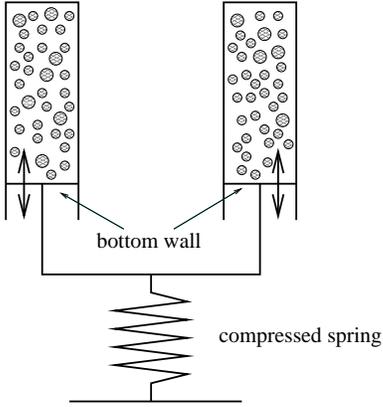} }}
\caption{\label{fig:b_spring} Schematic representation of Couette cell
  in vertical cross-section with movable bottom wall: possible
  realization of constant pressure experiment. }
\end{figure}
In this section we describe the simulations of sheared granular flow
in a system where the pressure on one of the walls is controlled at
all times. We refer to these boundary conditions as {\em controlled
pressure simulations}. When pressure is kept constant we have the case
of {\em constant pressure (CP)simulations}. In laboratory experiments
the CP condition can be realized not only by setting the constant
weight load on a movable top wall of a Couette cell\cite{miller96},
but also by setting the movable wall, for example, a bottom wall, on a
compressed spring, see Fig.~\ref{fig:b_spring} and
Ref.~\onlinecite{daniels03}. Any increase of the pressure inside the
cell results in an increase of the volume of the cell and vice
versa. The compressed spring realization is especially relevant if the
experiment is conducted in zero gravity.

In numerical simulations we adopt the following algorithm (which is an 
approximation of a compressed string) to control the pressure. 
The whole timespan of simulation is divided into small time
intervals $\Delta \tau$. At the beginning of each time interval we
calculate the total stress on the bottom wall, $\sigma$, using the
impacts of the particles during the previous time interval. More
precisely, we use Eq.~(\ref{eq:fce_n}) with $A_s=A_{base}$ and put
$\Delta t = \Delta \tau$. We then compute the vertical velocity of the
bottom wall, $v_b$, as follows:
\begin{equation}
  v_b = \left\{ \begin{array}{ll}
- {\bar v}_b, & \sigma > {\bar \sigma} + \delta \sigma / 2 \\
0           , & {\bar \sigma} - \delta \sigma / 2 < \sigma < {\bar
  \sigma} + \delta \sigma / 2 \\
+ {\bar v}_b, & \sigma < {\bar \sigma} - \delta \sigma / 2
\end{array}
\right.
\label{eq:vb}
\end{equation}
where ${\bar \sigma}$ and $\delta \sigma$ are the prescribed stress
and stress tolerance, and ${\bar v}_b$ is a prescribed maximum
instantaneous velocity of the bottom wall. Within each time interval
$\Delta \tau$ we run the controlled volume simulations with constant
velocity of the bottom wall given by Eq.~(\ref{eq:vb}). The resulting
changes of the volume of the cell adjust the stress back to the
prescribed value. This model leads to vibration of the bottom wall in
steady state; similar vibrations are also observed in
experiments.\cite{savage84b}

We consider the system without oscillations and with glued particles
on the top wall, Fig.~\ref{fig:gb}(a), as a basis for all our CP
simulations. We set the constant stress on the bottom wall being equal
to the average stress $\av{\sigma_n}$ on this wall which we measured
in CV simulations with $V=10\, rad/sec$.
The choice of the parameters used in the model (\ref{eq:vb}) is
\begin{eqnarray}
  {\bar \sigma} = 222 \, \rho_s \frac{\av{d}^2}{sec^2} \label{eq:p2-1} \\
  \delta \sigma = 20 \, \rho_s \frac{\av{d}^2}{sec^2} \nonumber \\
  \Delta \tau =T_w/25 \nonumber \\
  {\bar v}_b = \av{d}/\Delta \tau /1000 = 0.915 \av{d}/sec
  \label{eq:p2}
\end{eqnarray}

\begin{figure*}[!ht]
\centerline{\hbox{
\includegraphics{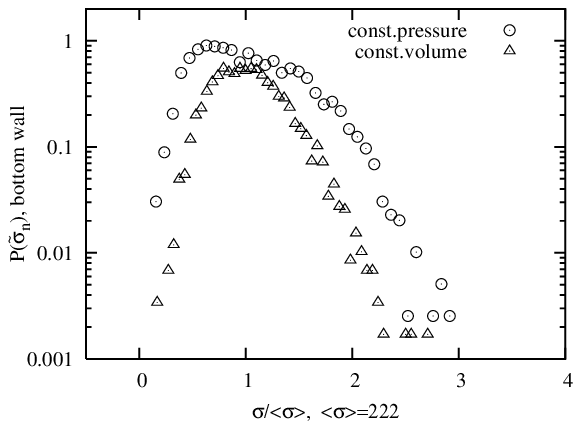}
\includegraphics{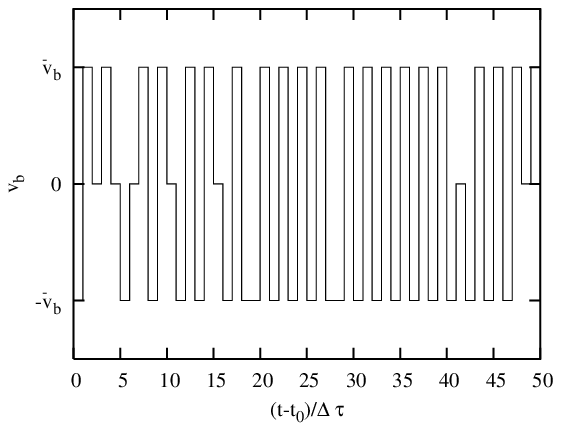}
}}
\centerline{\hbox{\hspace{0.1in} (a) \hspace{2.4in} (b) }}
\caption{\label{fig:f_base_02} (a) Stress on the base PDF for the case
  of CP simulations (circles), compared to CV results
  (triangles). Here $\Delta t = \Delta \tau = T_w/25$ and
  $A_s=A_{base}$ for both sets of results. (b) The velocity of the
  bottom wall in the CP simulations recorded after some late time
  $t_0$ in a steady state during 50 $\Delta \tau$ intervals. }
\end{figure*}
Figure~\ref{fig:f_base_02}(a) shows the PDF for the stresses on the
complete bottom wall in the case of CP simulations (circles) using
parameters given by (\ref{eq:p2-1})-(\ref{eq:p2}) and $\Delta t =
\Delta \tau $. For comparison, the results of CV simulations for the
same $A_s$ and $\Delta t$ are shown on the same plot (triangles). We
see that the CP distribution is wider compared to CV case. To
understand this difference it is useful to distinguish three sets of
collected stress data, each set containing the data from one of three
phases of the bottom wall motion: moving up, staying put, and moving
down (this separation of the stresses is possible only because $\Delta
t = \Delta \tau$, {\em i.e.} during the averaging time the bottom wall
does not change its velocity). In each phase we have controlled volume
evolution of the system with a different mean stress: highest in the
phase of the bottom wall moving up and lowest in the phase of the
bottom wall moving down. The resulting CP distribution must correspond
to the superposition of the distributions from three phases. The
degree of widening of the stress distribution can be estimated using
the following argument.

If $\Delta v_{y}^{(CV)}$ is a change in the normal component of
velocity of a particle interacting with a stationary sensor and
$\Delta v_{y}^{(CP)}$ is a change in the normal component of the
velocity of the same particle interacting with a sensor that has the
velocity $v_b$, then $\Delta v_y^{(CP)} =\Delta v_y^{(CV)} + v_b$
(note that $v_b$ can take three different values, see
(\ref{eq:vb})). If we have $N$ particles of average diameter $\av{d}$
interacting with a sensor $A_s$ during $\Delta t$ then, using the
definition (\ref{eq:fce_n}),
\begin{equation}
  \sigma_n^{(CP)} = \sigma_n^{(CV)} + K_1 \,\frac{N v_b}{A_s \Delta t} 
\end{equation}
where $\sigma_n^{(CP)}$ is a stress on a moving sensor,
$\sigma_n^{(CV)}$ is a stress on a stationary sensor, and $K_1=
\frac{1}{6} \pi \rho_s \av{d}^3$. Therefore, one expect that the
distribution of $\sigma_n^{(CP)}$ is approximately $2K_1 \,\frac{N
v_b}{A_s \Delta t}$ wider then the distribution of $\sigma_n^{(CV)}$.
This argument predicts that $(width^{CP}-width^{CV})/width^{CV}\sim
0.1-0.2$, consistently with the results shown in
Fig.~\ref{fig:f_base_02}(a) (in the estimate we use $N=10$, based on
the results shown in Fig.~\ref{fig:rms1}(c)).  We note that the above
argument applies only to the normal stresses, since this is the
direction of $v_b$.

\begin{figure*}[!ht]
\hbox{
\includegraphics{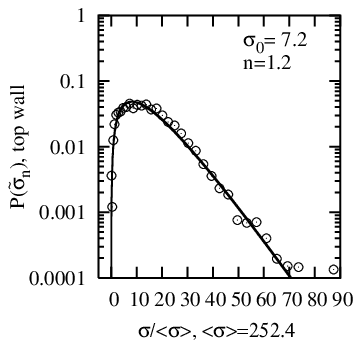} 
\includegraphics{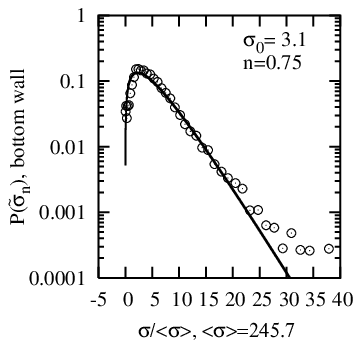} 
\includegraphics{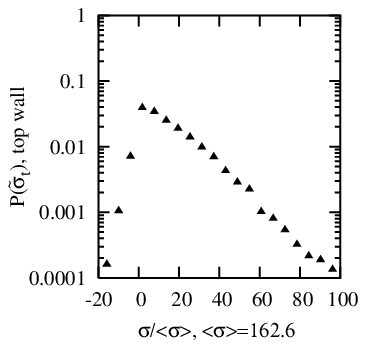} 
\includegraphics{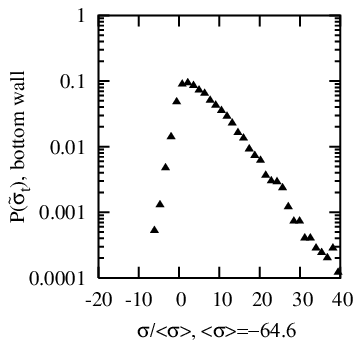} 
}
\abcd
\caption{\label{fig:f_03} Constant pressure simulations: Stresses for
the case without oscillations and with glued particles on the top
wall. Size of sensors $d_s=1.09 \av{d}$. (a) (c) - sensors are on the
top wall; (b) (d) - sensors are on the bottom wall; (a) (b) - normal
stress PDF; (c) (d) - tangential stress PDF. }
\end{figure*}

In principle, both normal and tangential stresses can also be affected
by the changes in the volume fraction and in the velocity profiles due
to the motion of the bottom wall. To estimate possible influence of
this motion, we note that it is characterized by the approximate
frequency of $1/( 2 \Delta \tau) = 457$ $Hz$ and the amplitude of
$\bar{v}_b\,\Delta \tau / 2 = 0.0005\, \av{d}$.  Despite high
frequencies, the vibrations of such a small amplitude are not expected
to significantly modify stress distributions, as also illustrated also
by the Fig.~\ref{fig:f_03}.

Figure~\ref{fig:f_03} shows stress PDF's on typical size sensors using
our typical averaging time $\Delta t = T_w/10$.
Comparing to the stresses in CV simulations,
Fig.~\ref{fig:forces_1}(a) and Fig.~\ref{fig:forces_2}(a), we do not
see any significant differences. This is because $\Delta t > \Delta
\tau$, therefore, during the averaging time the velocity of the bottom
wall can change once or twice, resulting in ``averaging out'' the
effect of the bottom wall motion. Therefore, we conclude that the
model specified by (\ref{eq:vb}-\ref{eq:p2}) is effective in keeping
constant stress, while not modifying significantly stress
distributions on the time scales of interest.

\subsection{\label{sec:bag} Bagnold scaling}
In this section we investigate the effect of the shearing velocity $V$
on the stresses. The relation between average normal stress and $V$
under certain conditions takes the form of {\em Bagnold
scaling}\cite{bagnold54}, where stresses increase with the square of
the shear rate. This effect has been studied in both CP and CV
settings.\cite{savage81,savage84b,campbell85,silbert01,bossis04} The
quadratic dependence of the stress on the shearing rate is observed
for the granular flows at high shearing rates or for lower volume
fractions, {\em i.e.} in the systems where particles are not involved
in multiple elastic deformations.\cite{campbell02}


\begin{figure*}[!ht]
\hbox{
\includegraphics{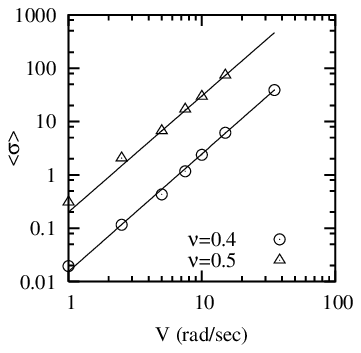} 
\includegraphics{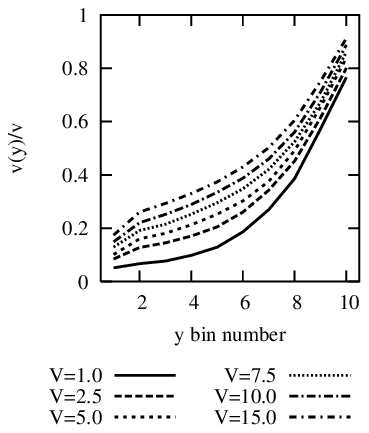}  
\includegraphics{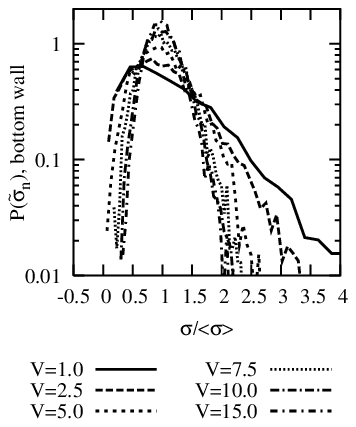} 
\includegraphics{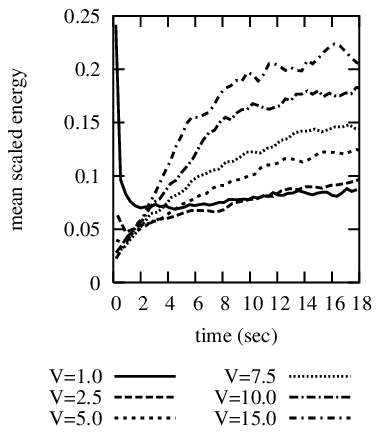}  
}
\abcd
\caption{\label{fig:cv} CV simulations. (a) Stress as a function of
  $V$ in log-log plot. Circles: $\nu =40\%$. Triangles: $\nu
  =50\%$. Solid lines are the best fits of power function as explained
  in the text. The rest of results shown is calculated using $\nu =
  40\%$. (b) Velocity profiles; (c) Stress PDF, bottom wall, $\Delta
  t=T_w/25$. Measurements are taken over last $\Delta T= 5000 \Delta t
  = 5.47 sec$ except for $V=1 rad/sec$, where $\Delta T = 2 sec$; (d)
  Scaled energy plots for different $V$'s.}
\end{figure*}

Figure~\ref{fig:cv}(a) shows log-log plot of the mean normal stresses
on the bottom wall versus $V$ at CV setting for the system shown in
Fig.~\ref{fig:gb}(a), {\em i.e.} with glued particles on the top wall
and no oscillations. For configurations of $\nu = 40\%$ and $\nu =
50\%$ these results confirm Bagnold scaling for the range of $V$'s
between 1 rad/sec and 35 rad/sec.  (Note that the $\nu = 50\%$ initial
state was obtained from the typical configuration by adjusting the
height of the cell according to (\ref{eq:h1})). The data are fitted by
the power function $\av{\sigma} = a V^b$ with $a=0.015$ and $b=2.20
\pm 0.2$ for $\nu = 40\%$ and $a=0.201$ and $b=2.18 \pm 0.2$ for $\nu
= 50\%$.

Figure ~\ref{fig:cv}(b) shows scaled velocity profiles for different
$V$'s in CV simulations. The profiles are higher for faster
shearing. The origin of this dependence may be related to the effect
of sidewalls on the sheared system. Indeed, when we shear the system
with elastic and smooth sidewalls, Fig.~\ref{fig:bnosh1}, the profiles
do not depend on $V$. Experimental studies\cite{losert00} also confirm
the independence of velocity profiles on $V$ in the case boundary
effects are significantly minimized. To explain these results we
consider the dependence of restitution parameters on the particle's
velocity in particle-sidewall interactions. The normal coefficient of
restitution (\ref{eq:restitution}) depends on the normal velocity of
colliding particle; however this velocity is not influenced
significantly by $V$. On the other hand, the coefficient of tangential
restitution, $\beta$, in the case of sliding contacts (\ref{eq:beta})
depends stronger on $V$ since it involves the ratio of normal to
tangential velocity, and the later scales with $V$. As it can be seen
from (\ref{eq:restitution}), lower ratio has the same effect on
dissipative mechanics as lower coefficient of friction,
$\mu$. Therefore, in the case of higher shear rates, the effect of
sidewalls on the granular flow is reduced, explaining the velocity
profiles in Fig.~\ref{fig:cv}(b).  

The rest of the results shown in Fig.~\ref{fig:cv} includes stresses
and mean scaled energies. We discuss them later in this Section in the
context of comparative study of constant volume and constant pressure
boundary conditions.
\begin{figure*}[!ht]
\centerline{\hbox{
\includegraphics{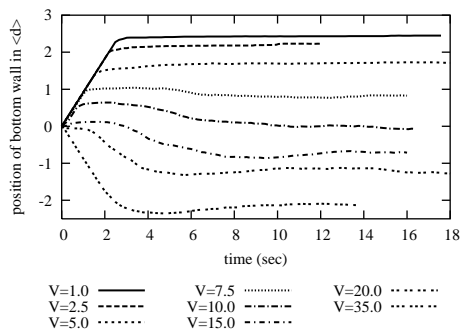} 
}}
\caption{\label{fig:cp_bw} Constant pressure simulations.  Position of
  the bottom wall, $h$, (in units of $\av{d}$) as a function of time
  for different shearing velocities.}
\end{figure*}

\begin{figure*}[!ht]
\hbox{
\includegraphics{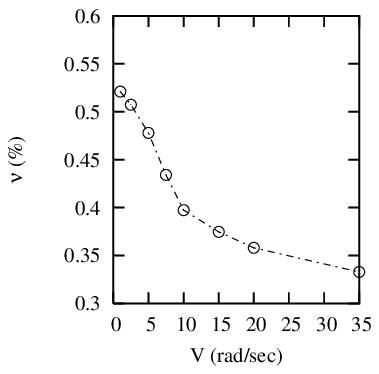} 
\includegraphics{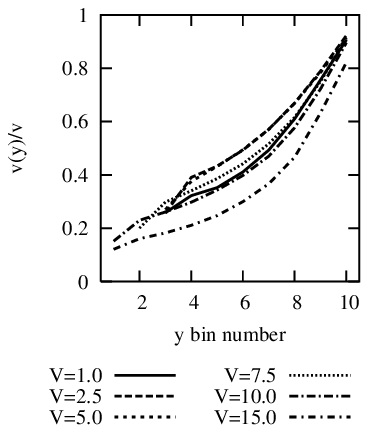}  
\includegraphics{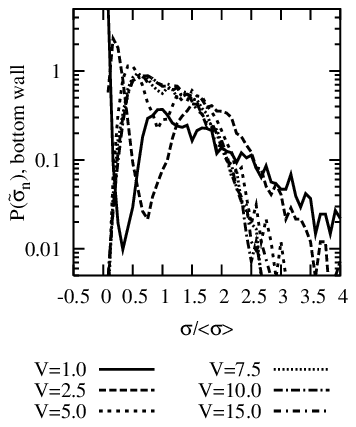}  
\includegraphics{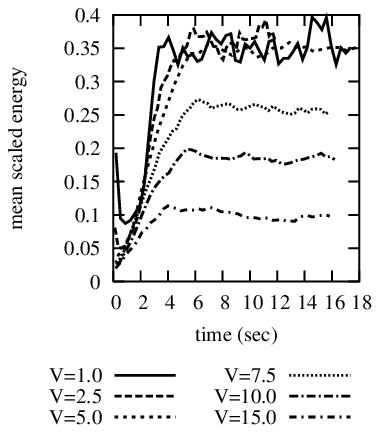}  
}
\abcd
\caption{\label{fig:bagnold01} CP simulations. (a) $\nu$ as a function
  of $V$; (b) velocity profiles; (c) Stress PDF, bottom wall. These
  distributions are obtained from the stress data taken every $\Delta
  t = \Delta \tau = T_w/25$ seconds over the last $\Delta T = 5000
  \Delta t $; (d) scaled energy plots.}
\end{figure*}

Next, we compare the effect of shearing velocity on the stresses for
CV and CP configurations. In CP simulations we choose the fixed value
of the stress on the bottom wall given by
(\ref{eq:p2-1}). Figure~\ref{fig:cp_bw} shows the vertical
position of the bottom wall $h$ as a function of time for different
$V$ ($h=0$ corresponds to the distance from the top wall equal to
$10.54 \av{d}$).
As expected, we see that slower/faster shearing leads to
smaller/larger stress, and to an adjustment of the wall position
occurring over some time interval (which is influenced by the ``spring'' 
parameters given by (\ref{eq:p2-1}-\ref{eq:p2})). The stabilized 
(long time) wall positions are used to calculate $\nu$ for each
$V$. Figure~\ref{fig:bagnold01}(a) shows resulting $\nu$'s.
We notice significant change in $\nu$ as $V$ is modified: $\nu$ at
high shearing (35 rad/sec) is about 1.6 times smaller then $\nu$ at
low shearing (1 rad/sec).


Figure~\ref{fig:bagnold01}(b) shows the effect of $V$ on the velocity
profiles. This result shows three types of different response to the
increase of $V$ in CP. For small $V$'s, $V \le 2.5 \, rad/sec$, the
shearing is stronger for larger $V$. For intermediate $V$'s, $2.5 \,
rad/sec \le V \le 5 \, rad/sec$ the velocity profiles are almost the
same, and for $V \ge 5 \, rad/sec$ the shearing decreases with an
increase of $V$. These results can be understood in terms of the
volume fraction, $\nu$, as a determining factor. For the smallest $V$
(solid line in Fig.~\ref{fig:bagnold01}(b)) $\nu$ is about $53\%$,
therefore local jammed areas can be formed inside the sample reducing
the overall mobility and shearing of the sample. These jammed areas
can manifest themselves as ``plateaus'' in velocity profile. One such
plateau can be seen in Fig.~\ref{fig:bagnold01}(b), solid line,
between $y$ numbers 4 and 5. As $\nu$ decreases with an increase of
$V$, the sample gets more fluidized and the shearing
improves. However, further increase of $V$ and decrease of $\nu$ makes
the system more ``compressible'': the dilation close to the top wall
is stronger, reducing the momentum flux from the top wall to the bulk
of the sample, resulting in weaker shearing. Therefore, the influence
of $V$ on a CP system is more complicated compared to a CV case, where
velocity profiles are changing monotonously with $V$, see
Fig.~\ref{fig:cv}(b).

The distributions of the normal stress on the bottom wall are shown in
Fig.~\ref{fig:cv}(c) for the CV and in Fig.~\ref{fig:bagnold01}(c) for
the CP. In both cases, CV and CP, the widths of distributions increase
with a decrease of $V$. The distributions differ in the following: in
CP case (Fig.~\ref{fig:bagnold01}(c)) the distribution shows a double
peak at low shearing around the prescribed value of the stress; This
double peak is absent in CV case. This double peak occurs since at low
$V$ the volume fraction is high, up to 52\%. Thus the collision rate
between the particles and the moving bottom wall is higher and, hence,
larger momentum is transfered to the granular system. If this momentum
is large enough, we have the situation in which the ``restoring
force'' of the bottom wall brings the system in just one interval
$\Delta \tau$ from the overstressed state to under-stressed state or
vice versa. Therefore, the velocity of wall, (\ref{eq:vb}), in steady
state alternates its value between positive and negative without
taking zero value, resulting in two peaks in stress distribution: one
for overstressed state and another one for under-stressed state.

Finally, Fig.~\ref{fig:bagnold01}(d) shows the time evolution of the
mean scaled energy of the system. This figure shows that for large
$V$'s, the equilibrating time decreases with an increase of $V$, while
for low shearing this trend is opposite. This effect is due to the
fact that the equilibrating time depends on the collision rate, which
in turn is the function of $V$ and $\nu$. At high shearing $\nu$
changes slowly with shearing, see Fig.~\ref{fig:bagnold01}(a), so
mostly $V$ determines the collision rate, leading to the observed
decrease of the equilibrating time with faster shearing. On the other
hand, at slow shearing, small changes in $V$ cause large changes in
$\nu$. Here mostly $\nu$ determines the collision rate, leading to the
observed increase of the equilibrating time with faster shearing.

%% file: base_conclusions.tex
\section{Conclusions}
The presented studies concentrate on 3D event driven simulations in
the Couette geometry with top rotating wall and with physical boundary
conditions in zero gravity. The velocity and volume fraction
distributions are strongly dependent on various boundary conditions,
such as: top shearing wall properties, side wall properties, presence
of oscillations of the bottom wall, or intensity of shearing. The
overall volume fraction of studied granular system is 40 \%, however,
the shearing and vibrating walls impose the non-uniformity in volume
fraction distribution with the formation of high-dense band (cluster)
where local volume fraction reaches up to 60\%. The cluster can
respond to shearing in different ways. For example, while very rough
and inelastic top wall, but without glued particles, cannot induce any
significant shearing when side-walls are dissipating, it imposes a
shear and high tangential velocities of all particles in the case of
smooth and elastic side-walls.
Typically, inside the cluster the shear rate is very small. However,
the presence of the glued particles on the top wall considerably
increases shear rate inside this cluster.

The presence of oscillations of the bottom wall seems to have three
effects: 1) the slippage velocity at the bottom increases. 2) the
dense cluster is located further away from the bottom wall compared to
the systems without oscillations. 3) equilibrating times to reach a
steady state are shorter in the systems with oscillations due to the
increased collision rate.

The equilibrating dynamics involves attaining the steady state values
of volume fractions and velocities. The time scales of these two
equilibrating processes are different: in most cases, steady state
profile in volume fraction is attained very fast, while velocity
profiles reach their final shape much later. However, the volume
fraction distribution takes longer to reach a steady state when the
rate of energy input due to shearing is low. This occurs in the
systems without glued particles and without oscillations. For example,
in so called ``delayed dynamics'' regime, described in
Sec.~\ref{sec:nosh}, the time needed for the cluster to accumulate
enough energy to dis-attach from the bottom wall can be very
long. However, once this happens, the velocity profile changes
drastically. More generally, these simulations show that the
velocities and volume fractions strongly depend on boundary
conditions: simulations using {\em e.g.} periodic boundary conditions
may miss a number of interesting effects presented in this work.

We have also analyzed stress and stress fluctuations on the
boundaries.
We find that the distribution of normal stresses in a system of 40\%
volume fraction and rapid granular flow can be described by the same
functional form (\ref{eq:fdist}) as used for stress distribution in
static dense system, {\em i.e.}  with power-law increase for small
stresses and exponential decrease for large ones. Considering this
functional form as a useful guide, we predict the behavior of main
characteristics of stress distributions as a function of sensor area
and averaging time. These predictions describe well the width of
distributions for small sensors and short $\Delta t$'s. However, they
fail for large sensors and long averaging times, suggesting the
existence of additional correlations not accounted by our functional
form. The correlations may be related to the existence of ``denser
structures'' in the simulated systems.

We simulate ``constant pressure'' boundary condition and we discuss
the differences between the results in constant volume and constant
pressure settings. A key observation here is very different reaction
to an increase of shearing velocity in the system with constant
pressure boundary condition compared to the systems with constant
volume boundary condition. In the first case the shearing velocities
are increasing because of the velocity dependence of the coefficients
of restitution while in the second case they are initially increasing
and then decreasing because of the changes of volume fraction with
imposed shear.

\begin{acknowledgments}
We acknowledge the support of NASA, grant number NNC04GA98G. We thank
Robert P.  Behringer, Karen E. Daniels, and Allen Wilkinson for useful
discussions.
\end{acknowledgments}
